\documentclass[aps,prl,floatfix,twocolumn,nofitinbib,hyperref=pdftex,citeautoscript]{revtex4}
\usepackage{epsfig}

\usepackage{amsmath,bm}
\usepackage{amssymb}
\usepackage{braket}
\usepackage{amsfonts}
\usepackage{dsfont}
\usepackage{comment,color}
\usepackage{lipsum}
\usepackage{hyperref}

\hypersetup{colorlinks=true,linkcolor=blue,anchorcolor=blue,citecolor=blue,filecolor=blue,urlcolor=blue,bookmarksnumbered=true,pdfview=FitB}

\usepackage[dvipsnames]{xcolor}

\begin{document}

\title{Non-Hermitian higher-Order Weyl semimetal with surface diabolic points}

\author{Subhajyoti Bid}
\author{Gaurab Kumar Dash}
\author{Manisha Thakurathi}

\affiliation{Department of Physics, Indian Institute of Technology Delhi, Hauz Khas, New Delhi, India 110016}

\date{\today}

\begin{abstract}
Higher-order topology in non-Hermitian (NH) systems has recently become one of the most promising and rapidly developing fields in condensed matter physics. Many distinct phases that were not present in the Hermitian equivalents are shown in these systems. In this work, we examine how higher-order Weyl semimetals are impacted by NH perturbation. We identify a new type of topological semimetal, i.e., non-Hermitian higher-order Weyl semimetal (NHHOWS) with surface diabolic points. 
 We demonstrate that in such an NHHOWS, new exceptional points  inside the bulk can be created and annihilated, therefore allowing us to manipulate their number. At the boundary, these exceptional points are connected through unique surface states with diabolic points and hinge states.  
For specific system parameters, the surface of NHHOWS behaves as a Dirac phase with linear dispersion or a Luttinger phase with a quadratic dispersion, thus paving a way for Dirac-Luttinger switching. Finally, we employ the biorthogonal technique to reinstate the standard bulk boundary correspondence for NH systems and compute the topological invariants. The obtained quantized biorthogonal Chern number and quadruple moment topologically protect the unique surface and hinge states, respectively.
\end{abstract}

\maketitle

\textit{Introduction.\textemdash}
 In the past decade, topological phases such as topological insulators \cite{su1979solitons,asboth2016schrieffer,bid2022topological, Frank,Dsen,sato2017topological,Deb_TI}, topological superconductors \cite{Arijit_TS, MT_Kitaev,Tanay_sc,MT_Majorana,JK1,perez2018ssh,kitaev2001unpaired,beenakker2013search,JK3,Silas} and Dirac/Weyl semimetal \cite{Anton1, Arijit_weyl, Anton2, MT_cdw, MT_weyl} in Hermitian systems have been attracting substantial attention.
 Currently, the field of NH topological systems \cite{Nori,huges,zhen,ashida2020non,Sasa,Carl, Soori} is also rapidly emerging with a variety of potential applications in condensed matter physics. One of the key features of NH systems is the existence of a unique branch point in the  spectrum, known as an exceptional point (EP) where both eigenenergies, as well as eigenvectors, coalesce. This is quite different from a conventional well-known degenerate point in Hermitian systems, named as diabolic point (DP), at
which only the eigenenergies coalesce. Although DPs have a variety of characteristics,  their physics in non-Hermitian systems is less explored. It has been observed generally that the DPs are not stable and split into EPs or evolve into exceptional rings in the presence of NH terms in the Hamiltonian \cite{Duan,Nori2}. 

 In this Letter, we describe a method for generating DPs on the surface of an NH system while sustaining  EPs in the bulk of the systems. We illustrate that the DPs change their location on the surface as well as dispersion around them when the system parameters vary. This allows us to achieve the Dirac to Luttinger phase switch on the surface. Notably, DPs emerge precisely at those locations in momenta where the Dirac nodes are formed in the higher order Dirac semimetal (HODS) \cite{huges,wang,lin2018topological} thus retaining the memory of HODS phase. We begin with a $C_4^z$ symmetric HOWS which is created by breaking the degeneracy of each Dirac node in HODS. We unravel the fate of such a system under NH Weyl perturbation  (responsible for breaking the degeneracy of Dirac points) due to nonreciprocal couplings in the system. The NH perturbation in the Hamiltonian  breaks each Weyl node into two EPs. Starting from eight we can decrease their number to six, four, two, and zero by annihilating them or vice versa by tuning the system parameters. Some of the EPs are connected either by bulk Fermi arcs (FA) \cite{Bulk_FA} or surface FAs when the system is finite along one axis. Remarkably, the energy eigenvalues of the surface FAs, have degenerate DPs and a variety of dispersion spectra around them. For certain system parameters, we can achieve a Dirac phase having an absolute value of energy $ \sim |k|$  as well as a  Luttinger phase with $ k^2$ dispersion relations. Since these EPs are connected by surface states so we identify them as normal-order EPs which are characterized by quantized open boundary Chern number. When the system is finite along two directions, we find that the innermost EPs closest to the $\Gamma$ point in BZ are connected by hinge states, which are completely flat bands localized only at the hinges of the system. In recent literature, hinges are characterized by higher-order topology, and therefore the corresponding EPs connecting them are of higher order in nature \cite{hodaei2017enhanced,mandal2021symmetry}. Among many, one of the interesting feature of NH systems is the breaking down of  bulk boundary correspondence (BBC). 
This is due to the fact that these systems exhibit Non-Hermitian skin effect (NHSE) \cite{okuma2020topological,kawaexcep,zhang2021observation}. To restore BBC, generally, two distinctive approaches are used, (i) generalized Brillouin zone approach \cite{genbbc} and (ii) biorthogonal real space approach \cite{song2019non}. Due to the simplicity and convenience, we use the latter approach to calculate the topological invariants, namely open boundary Chern number \cite{song2019non} and quadrupole moment \cite{benalelectric,multipole}, for unraveling the theory of normal as well as higher-order topological phase of this novel semimetal. Our theoretical proposal can readily be realized on topolectrical circuit lattices \cite{topoelectric}. Consequently, it may be foundational for future experiments
aimed at a controlled manipulation of Dirac to Luttinger dispersion.

\textit{Model and formalism.\textemdash}
We consider HOWS, constructed by stacking 2D quadrupole insulators along the $z$-axis  \cite{huges,multipole}. The Hamiltonian using spinless fermion has the form $H_{\text{NHW}}(k)=H_0(k)+i\, \delta\sigma_0\kappa_1$, where $H_0(k)$ is the Hermitian part and is given by,
\begin{align}
    &H_0(k)=  \sum_{j=1}^4 h_j \Gamma_j +  m \sigma_0\kappa_2.
    \label{HNHW}
\end{align}
Here $h_{1/3}=\sin k_{y/x}$ and $h_{2/4}=(\gamma +0.5\cos k_z+ \cos k_{y/x})$. The intercell hopping is denoted by $\gamma$, the direct product of Pauli matrices have the form $\Gamma_0=\sigma^3\kappa^0$, $\Gamma_i=-\sigma^2\kappa^i$,  $\Gamma_4=\sigma^1\kappa^0$, and $m$ is intracell coupling with $\delta$ added as the NH contribution to it. In Eq. (\ref{HNHW}), spinless time-reversal symmetry $\mathcal{T}=\mathcal{K}$, where $\mathcal{K}$ is complex conjugate, mirror symmetries ($M_x$ and $M_y$) are broken due to the $m$ term.  However, it preserves $C^z_4$, inversion ($\mathcal{P}$), $M_x\mathcal{T}$ and $M_y \mathcal{T}$. This term splits the Dirac nodes into two Weyl nodes with opposite monopole charges. It has been shown in Ref. \cite{huges} that utmost four Weyl nodes of both first-  and second-order connected via surface and hinge arcs respectively, can be obtained. Furthermore, with the addition of the non-hermitian coupling $\delta$, multiple EPs emerge where both real, as well as complex part of the energy, goes to zero. The $\delta$ term preserves charge conjugation ($\mathcal{C}$) and $\mathcal{C}\mathcal{P}$. 
 However, it breaks  symmetries like  inversion ($\mathcal{P}$), $C^z_4$, $M_x\mathcal{T}$ and $M_y \mathcal{T}$, chiral, $\mathcal{P}\mathcal{T}$ \cite{natptsym}, anti $\mathcal{P}\mathcal{T}$, and reciprocity $\mathcal{R}$. 

\begin{figure}[t]
    \centering
    \begin{tabular}{cc}
   {\includegraphics[width=0.45\linewidth]{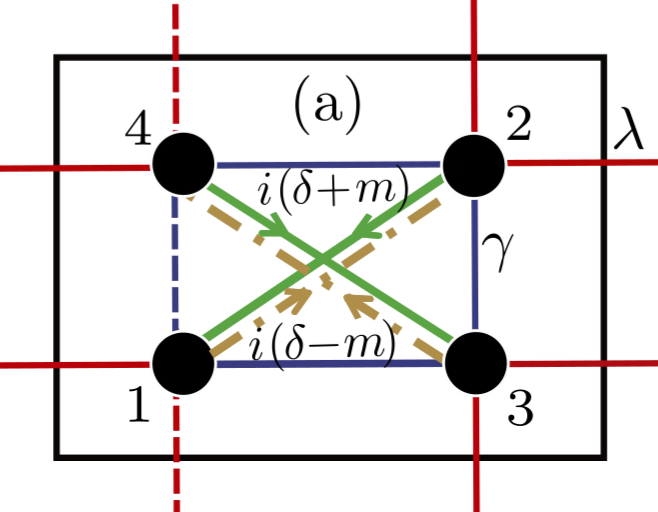}}& 
   \hspace*{-0.6cm}
   \raisebox{-0.5cm}
   {\includegraphics[width=0.51\linewidth]{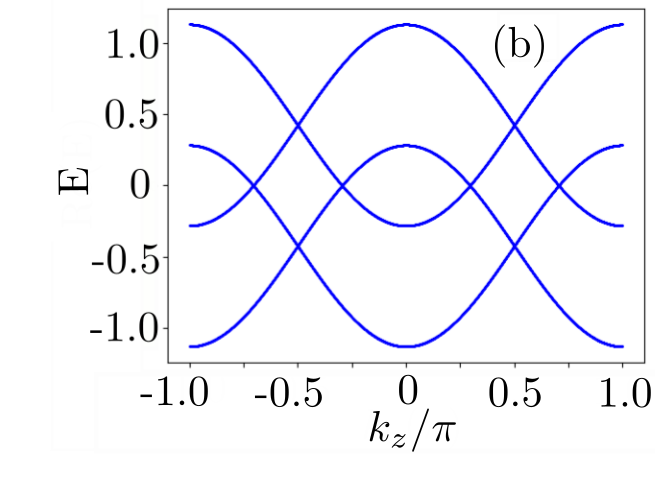}}\\
    \includegraphics[width=0.51\linewidth]{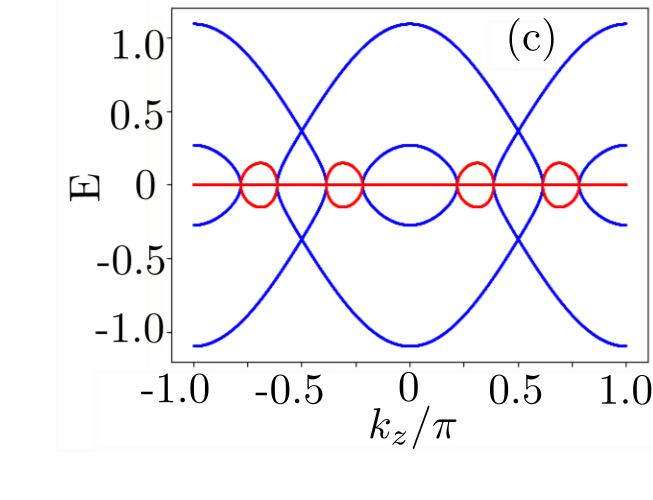}&
   \hspace*{-0.2cm}
   \raisebox{0.3cm}
   {\includegraphics[width=0.48\linewidth]{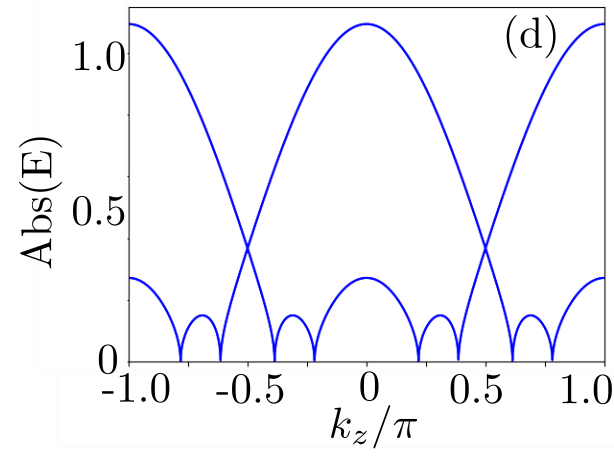}}
   \end{tabular}
    \caption{(a) Schematic diagram of unit cell with intercell hopping as $\lambda$, nearest neighbor intracell hopping as $\gamma$ and NH next nearest neighbor intracell hopping as $i(\delta\pm m)$. Dashed lines signify the opposite sign of hopping compared to solid ones. The bulk energy spectrum of (b) $H_0(k)$ with four DPs for $\gamma=-1$, $m=0.6/\sqrt{2}$, and $\delta=0$ and (c) for $H_{\text{NHW}}(k)$ having eight EPs  when 
 $\delta=m/2$ with blue (red) representing the real (imaginary) part of energy eigenvalues.  The non-Hermitian term, $\delta$ creates new EPs in the system, therefore the number of EPs changes as we vary the value of $\gamma$ and $\delta$, see \cite{SM}.  (d) Plot of the absolute energy as a function of $k_z$ with eight EPs for the same system parameters. }
    \label{Energy_Spectra}
\end{figure}

The energy spectrum of $H_{\text{NHW}}(k)$ is $E^2_{\pm}(k)=a(k) \pm 2 \sqrt{b(k)}$, where $a(k)=\sum_{i=1}^4 h_i^2 + m^2 - \delta^2$ and $b(k)=(h_2^2 + h_4^2) (m^2-\delta^2) + 2 i h_1 h_2 m \delta$. In Figs. \ref{Energy_Spectra} (b) [(c) and (d)], we show the spectrum of $H_0(k)$ [$H_{\text{NHW}}(k)$] as a function $k_z$, with a set of parameters, producing four Weyl nodes [eight EPs].   We have confirmed numerically that the EPs are formed only at $k_x=0$, therefore from now on we work with this parameter with the highest number of EPs in the bulk. For a detailed discussion of the EPs phase diagram, see \cite{SM}. 
\begin{figure}[b]
    \centering
    \includegraphics[width=0.48\linewidth]{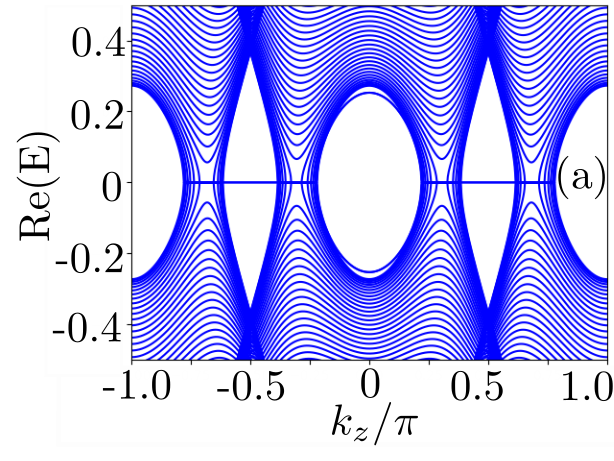}
    \includegraphics[width=0.48\linewidth]{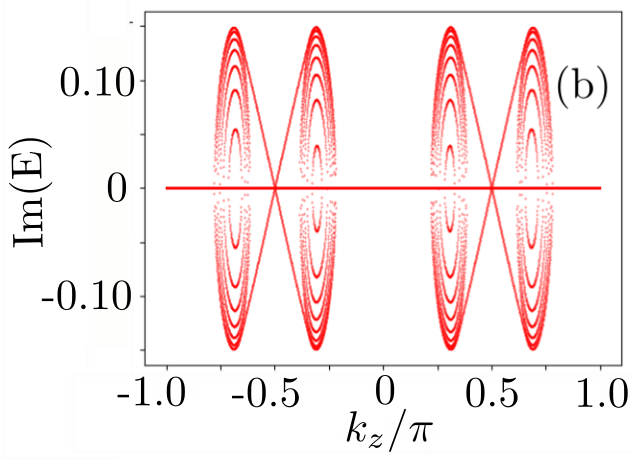}
    \includegraphics[width=0.48\linewidth]{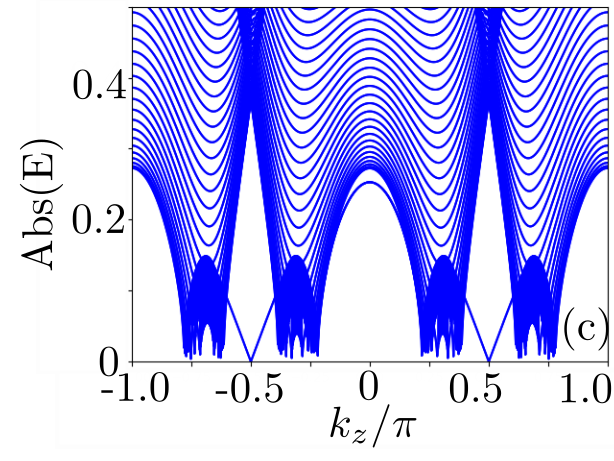}
    \hspace*{0.05cm}\includegraphics[width=0.5\linewidth]{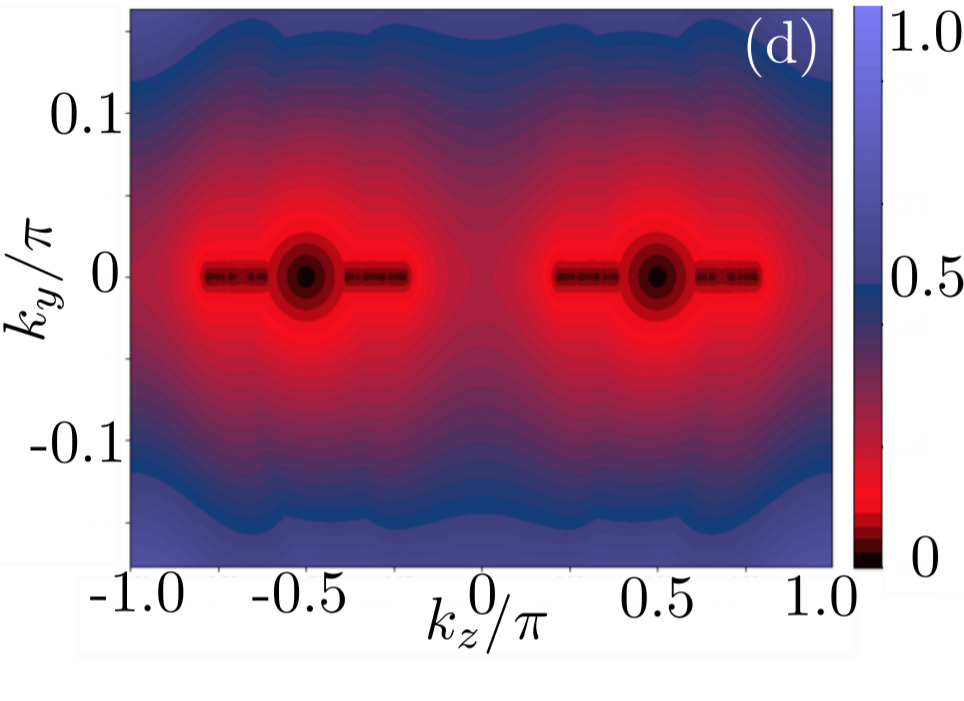}
    \includegraphics[width=0.48\linewidth]{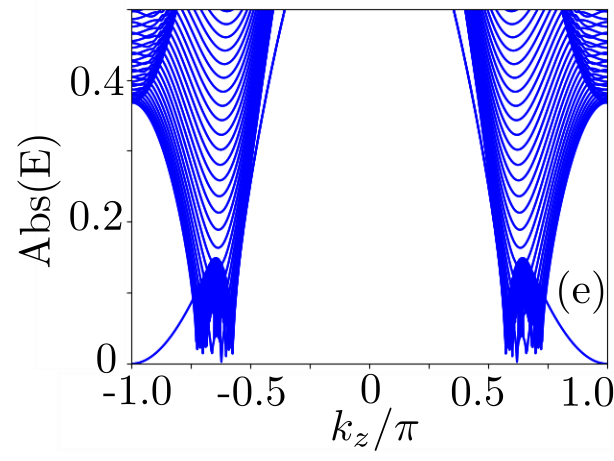}
    \includegraphics[width=0.48\linewidth]{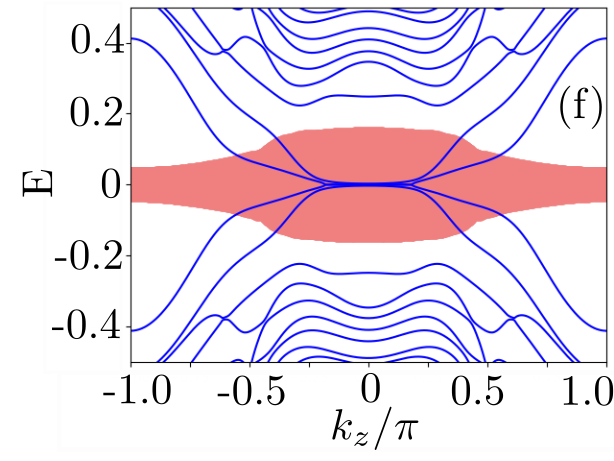}
    
    \caption{ (a) Real part of energy spectrum of $H_\text{NHW}$ for $x$ surface (at $k_y=0$) as a function of $k_z$ with same parameters as in Fig. \ref{Energy_Spectra}. Out of eight EPs four are connected by surface Fermi arcs, thus identifying them as normal-order EPs. The remaining four are connected through bulk Fermi arcs.  (b) the corresponding imaginary part of the energy spectrum, forming a ring-like structure around EPs. Absolute energy spectrum for the $x$ surface (c) demonstrating the formation of DPs at $k_z=\pm \pi/2$ and  $k_y=0$ with linear momenta dependency and (d) contour plot in $k_y$-$k_z$ plane shows the formation of two Dirac cones around $(k_y,k_z)=(0,\pm \pi/2)$. The concentric circles are projections of Dirac cones. (e) Absolute energy eigenspectrum for $\gamma=-0.5$, having a dispersion with quadratic momenta dependency at BZ boundary. Therefore,  behaving like a Luttinger phase on the surface of NHHOWS. (f) Finally, 
 the real (blue) and imaginary (red) part of the energy spectrum for hinge states formed when we have OBC along $x$ and $y$ directions. It clearly shows that innermost EPs close to $k_z=0$ are connected through hinge states, therefore named as higher-order EPs. We have verified our results for different values of $\gamma$ as well, see \cite{SM}.}
    \label{fig2}
\end{figure}
Further, we explore surface states by considering tight binding Hamiltonian along $x$-axis. 
In Fig. \ref{fig2}(a),  we label the EPs from the extreme left, the EPs are connected via bulk FAs from positions 1 to 2,  3 to 4, 5 to 6, and 7 to 8, however, from positions 2 to 3 and 6 to 7 via  surface FAs. We call these EPs normal-order as they are only connected through the surface states. Contrary to previous studies of NHWSM,  the absolute value of $E$ reveals that these Fermi arcs are dispersive in nature, see Figs. \ref{fig2} (c), (d), and (e). Without loss of generality, to simplify the analytical calculation, we consider $m=2\,\delta$ throughout the paper. The value of $k_z$  where the DPs are formed is written as $k_{z0}=\pm \arccos[-2(1+\gamma)]$ \cite{SM}. This matches exactly with the numerically obtained $k_{z0}=\pm\pi/2 \, (0)$ for $\gamma=-1  \,(-0.5)$. The distance between the two DPs increases (decreases)  with the corresponding change in $\gamma$. At $k_{z0}=\pm \pi/2$, the functional form of absolute energy for the surface state is  $\propto |k_z|$ \cite{SM}.  Hence the surface of NHHOWS is in the Dirac phase, hosting two Dirac cones at $k_{z0}=\pm \pi/2$.
Interestingly at a particular value of $\gamma=-0.5$ these DPs merge at BZ boundary $k_z=\pm \pi$, having absolute energy dispersion as $ \propto  k_z^2$ \cite{SM}. This kind of low energy dispersion along $k_z$ resembles a quadratic Luttinger spectrum at BZ boundary \cite{boettcher2016superconducting,ghorashi2018irradiated,mandal2021transport}. Therefore, by tuning $\gamma$, we switch from  Dirac  to the Luttinger phase on the surface of NHHOWS. Further increasing $\gamma$ above -0.5,  the DPs completely merge, thus, gapping out the energy spectrum. Also, we have analytically as well as numerically obtained that with the variation of $\gamma$, the dispersion is always linear along $k_y$ and becomes non-linear only along $k_z$ direction \cite{SM}.  The contour plot of surface states in Fig. \ref{fig2}(d) reveals the fact that surface states are actually a collection of FAs, and as a result, they form cone-like structures in the $(k_y-k_z)$ plane. 
Next, we explore the hinges of the system by applying OBC along the $x$ and $y$ directions, we observe that two innermost EPs closest to $k_z=0$ are connected by hinge states, see Fig. \ref{fig2}(f). Therefore, we identify them as, higher-order EPs. This phase is characterized by a topological invariant named as quadruple moment $q_{xy}$ \cite{benalelectric}, which we will discuss in the next section. The absolute energy band diagram for hinge states confirms that, unlike surface FAs, these are perfectly flat bands connecting innermost EPs. The degeneracy of hinge states is four and in the Hermitian counterpart, each one of the states is localized at all of the corners of the $x-y$ plane. However, for our NH system, these are localized in either two lower or two upper corners depending on whether we plot left $\ket{\psi_L}$ or right  $\ket{\psi_R}$ wavefunctions \cite{SM}. We identify this fascinating effect as higher order non-Hermitian skin effect (HONHSE). Moreover, we have looked at models with different symmetry classes and discovered that the peculiar surface feature and the hinge state behavior remain unchanged. This illustrates the stability of the NHHOWS with surface DPs, for more information, see \cite{SM}.\\

\begin{figure}[t]
    \centering
    \includegraphics[width=0.52\linewidth]{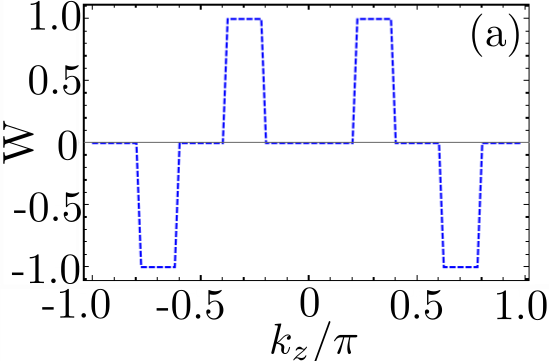}
    \includegraphics[width=0.5\linewidth]{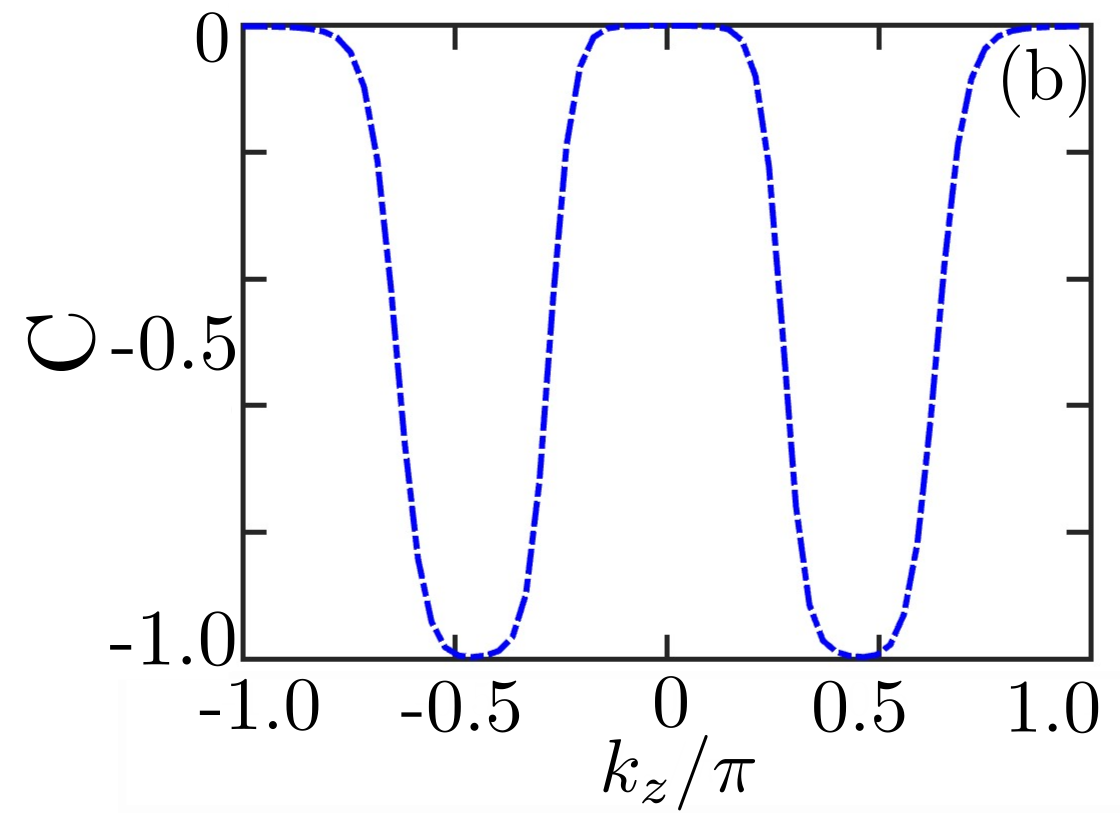}
    \includegraphics[width=0.48\linewidth]{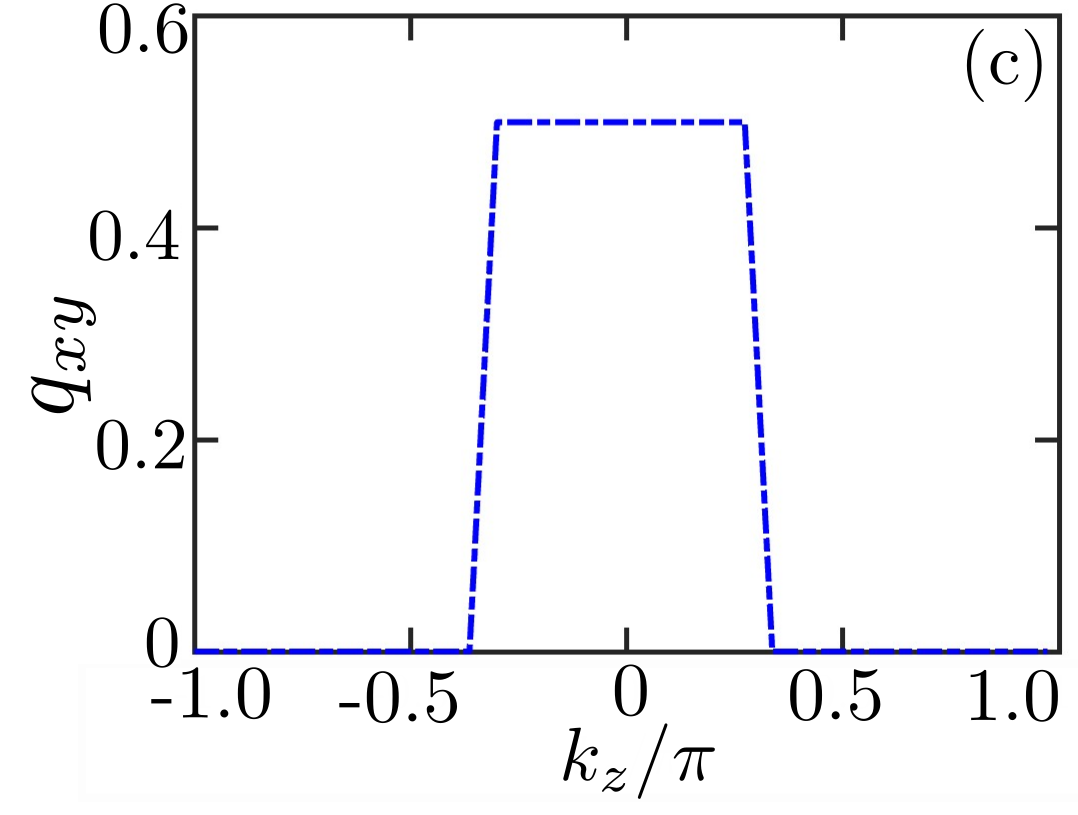}
    \caption{(a) Plot of one-dimensional winding number, $W=W_1+W_2$ as a function of $k_z$ at $k_x=0$.  It flips signs when it passes through EPs, highlighting their topological nature.
    (b) The open boundary  Chern number has been plotted as a function of $k_z$, it  takes an integer value of -1 only in the region where EPs are connected through surface states (d) higher-order topological invariant quadruple moment $q_{xy}$ is quantized to a value of half where hinge state connects the innermost EPs. The system parameters are the same as those used in Fig. 2 (a)-(d) and (f) for the Dirac phase. However, we have verified that all three invariants work perfectly for the Luttinger phase as well.}
    \label{topo}
\end{figure}

\textit{Topological invariants\textemdash}
In order to characterize the topology of these bulk EPs, we first adopt the formulation of 1D winding number from a previous study \cite{nhhigherweyl}.  We modify the Hamiltonian's basis in Eq.(\ref{HNHW}) through the $y$-axis rotation of the Pauli matrices ($\sigma^i$ and $\kappa^i$). The transformations are  $\sigma^x \rightarrow \kappa^y\sigma^x$, $\sigma^y \rightarrow \sigma^y$, and $\sigma^z \rightarrow \kappa^y\sigma^z$ followed by  $\kappa^x \rightarrow \sigma^y\kappa^x$, $\kappa^y \rightarrow \kappa^y$, and $\kappa^z \rightarrow \sigma^y\kappa^z$. This converts the Hamiltonian to an off-block diagonal at $k_x=0$ having lower (upper) block as $Q_{1}$ ($Q_2$) along with the property $Q_2 \neq Q_1^{\dagger}$ for NH systems. Moreover, these transformations reveal Hamiltonian's sublattice symmetry \cite{kawabata2019symmetry}. The Winding number as a function of $k_z$ at $k_x=0$ is given by the equation,
\begin{equation}
    W_{1/2}(k_z)=\int_{-\pi}^{\pi}\frac{1}{2\pi i}\partial_{k_y} \big[\log\big(\text{det}[ Q_{1/2}(k_y,k_z)]\big)\big]dk_y.
\end{equation}
The topological nature of the EPs is demonstrated by the fact that $W=(W_1+W_2)$ flips sign and jumps between 0 and $\pm 1$ as it crosses any of the EPs, see Fig. \ref{topo}(a). We have already illustrated how the NH skin effect significantly affects our model in both normal and higher-order phases \cite{okuma2020topological,zhang2021observation}. Therefore, instead of using the bulk Hamiltonian, we employ a biorthogonal method  to restore BBC in NH systems.  
  We compute the biorthogonal real space Chern number as a function of $k_z$ for Hamiltonians having OBC along two axes. Its  quantized value guarantees the topological characteristics of the surface states connecting the EPs. 
We work with the OBC eigenbasis. For NH systems, one has right and left eigenstates satisfying the eigenvalue equation as,  $H\ket{nR}=E_n\ket{nR}$ and $H^{\dagger}\ket{nL}=E_n^*\ket{nL}$, respectively. The left and right eigenstates are chosen in such a way that the biorthonormality is maintained, $\braket{mL|nR}=\delta_{mn}$, $\braket{mL|n'R}=\braket{m'L|nR}=0$. The primed states  are  the chiral partners of right (left) with negative eigenenergy $-E_n$ ($-E_n^*$). Furthermore, we compute the bulk band projection operator, $\hat{P_{\alpha}}=\sum_{n \in \alpha}\ket{nR}\bra{nL}$. The sum is over all the unoccupied bands (labeled as $\alpha$) i.e. below the Fermi level. Using $\hat{P_{\alpha}}$, the open boundary Chern number is calculated as,
\begin{equation}
    C_\alpha=\frac{2\pi i}{l_{x}'l_{y}'}Tr'\left(\hat{P_\alpha}[[\hat{X},\hat{P_\alpha}],[\hat{Y},\hat{P_\alpha}]]\right).
\end{equation}
$\hat{X}$ and $\hat{Y}$ are coordinate operators along $x$ and $y$  axes for a 2D slice corresponding to each value of $k_z$, defined as $\hat{X}_{m m'}=x\delta_{m m'}$ and similarly for $\hat{Y}_{n n'}=y \delta_{n n'}$ with $1\le x\le l_x$ ($1\le y\le l_y$) where $l_x (l_y)$ is the size of the system along $x (y)$ with unit lattice spacing. Also, $l_{x/y}'$ is defined as $l_{x/y}'=l_{x/y}-2l_0$ where $l_0$ is a boundary layer that is being removed from $l_{x/y}$. After excluding the boundary layer only bulk information is captured by $Tr'$ which is taken over the middle region. The open boundary Chern number for each 2D $k_z$ slice takes a quantized value, $C=-1$ exactly in the region where surface states connecting the EPs appear  otherwise its value is zero, see Fig. \ref{topo}(b). Thus, reflecting the topological character of the collection of surface FAs. Therefore, this approach gives us a powerful tool for computing the Chern number in real space and successfully capturing the topological nature of the surface states. 

As discussed earlier due to the presence of HONHSE these states are forced to  two of the corners rather than all four separately.  
We confirm the existence of  higher-order topology by computing the higher-order topological invariant that is the quadruple moment, $q_{xy}$ \cite{lin2018topological,multipole,benalelectric}. Motivated by the previous success of the biorthogonal approach while computing the Chern number, we reuse it to calculate the $q_{xy}$ using the formula written as \cite{wu2021floquet}

\begin{equation}
    q_{xy}=\left (\frac{\text{Im} [ \text{ln} \, (\text{det} \,\hat{Q}) \, ]}{2\pi}-\frac{\sum_i \hat{X}_i\hat{Y}_i}{2 l_x l_y}\right) \,\, mod \,\, 1,
\end{equation}
where, $\hat{Q}$ is a matrix whose elements are given by, $\hat{Q}_{m n}=\braket{mL|e^{\frac{2\pi i\hat{X}\hat{Y}}{l_x l_y}}|n R}$ and $\hat{X}$ and $\hat{Y}$ are the same coordinate operators used to calculate Chern number in the previous section as well. 
For each 2D $k_z$ slice from $[-\pi,\pi]$, we observe that $q_{xy}$ takes quantized value of half where hinge states appear (connecting the innermost higher-order EPs), see Fig. \ref{topo}(c). Therefore, again we are able to successfully characterize the higher-order topology of the hinge states.

\textit{Conclusions\textemdash} 
In summary, we have examined non-Hermitian HOWS and shown that NH term provides us a potent tool to  annihilate or create new EPs in the bulk of a system. Some of the EPs are connected through surface (hinge) states and are of normal order (higher-order). We identify this new type of topological semimetal as NHHOWS with surface diabolic points. The surface of such a semimetal features some exotic phases like  Dirac and Luttinger with linear and quadratic dispersion, respectively.  This allows Dirac to Luttinger phase switching on the surface of NHHOWS as a function of system parameters. Additionally, we describe the topological properties of the EPs by first computing a 1D winding number, then implementing the biorthogonal method to obtain the open boundary Chern number and $q_{xy}$. The latter two invariants identify the system's topological region, where surface and hinge states appear.
The Chern number has an integer value of -1 and $q_{xy}$ takes quantized value of half. Moreover, the biorthogonal technique stands out to be an essential tool in the context of computing topological invariants and subsequently capturing normal as well as higher-order topological phases of this system in a more versatile manner. 

\textit{Acknowledgments\textemdash}  For financial support, S.B. thanks CSIR, India and
M.T. thanks Science and Engineering Research Board (India) 
grant SRG/2022/001408 and Young Faculty Incentive Fellowship from IIT Delhi. The authors thank F. Song for stimulating
discussions on related topics.

\bibliography{sample.bib}

\newpage
\onecolumngrid
\begin{center}
{\bf \large{SUPPLEMENTARY MATERIAL: Non-Hermitian higher-Order Weyl semimetal with \\
\vspace*{0.1cm}
surface diabolic points}} \\
\vspace*{0.2cm}
Subhajyoti Bid, Gaurab Kumar Dash, Manisha Thakurathi \\
{\it Department of Physics, Indian Institute of Technology Delhi, Hauz Khas, New Delhi, India 110016}
\end{center}
\setcounter{equation}{0}
\setcounter{figure}{0}
\setcounter{table}{0}
\setcounter{page}{1}
\makeatletter
\renewcommand{\theequation}{S\arabic{equation}}
\renewcommand{\thefigure}{S\arabic{figure}}
\section{Phase space analysis for the number of EPs in bulk}
The NH Hamiltonian $H_{NHW}=H_0(k)+i\delta\sigma_0\kappa_1$ in the main text hosts two, four, six, and a maximum of eight EPs in the bulk. The parameters chosen in the main text are such that there are eight EPs in the bulk. The location of these EPs as the function of $k_z$ is obtained by solving the dispersion relation $E(k_z)=0$, where,
\normalsize{
\begin{equation}\label{eq1}
    E(k_z)=\pm \frac{1}{2} \sqrt{8 \gamma  (\gamma +2)-4 \delta ^2+8 (\gamma +1) \cos (k_z)\pm 4 \sqrt{\left(2 m^2-\delta
   ^2\right) (2 \gamma +\cos (k_z)+2)^2}+\cos (2 k_z)+4 m^2+9}.
\end{equation}
}

\begin{figure}[h]
    \centering
    \includegraphics[width=8cm]{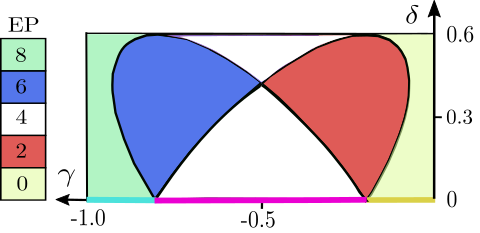}
    \caption{The phase diagram for the number of EPs in the bulk with elliptical phase boundaries. From the extreme left ($\gamma=-1$) we start with eight EPs and gradually annihilate them to decrease their number. One cannot directly reach from eight to zero EPs. There exist an interesting point in the phase diagram with values $\gamma=-0.5$ and $\delta \sim 0.4$, which we name the triple point of EPs in the bulk. Infinitesimal change around this point marks a phase transition and takes us to phases with different no. of EPs.}
    \label{phase}
\end{figure}

Here, we have fixed $k_x$ and $k_y$ to be at zero. The eight EPs are located at values,
$\pm [\pi-\arccos{\left(2 \gamma +\delta +m'+2\right)}$, $ \arccos \left(-2 \gamma +\delta \pm m'-2\right)$, $ \arccos \left(-2 \gamma -\delta + m'-2\right)]$ with $m'=\sqrt{2m^2-\delta^2}$. For a fixed value of $\delta$, when $\gamma$ varies such that one goes from a phase with eight EPs to six EPS, two of the $k_z$ values where EPs appear become complex (the argument of the arccos have a value greater than one). Similar behavior has been observed whenever we cross the phase boundaries in Fig. \ref{phase}.
Their number clearly depends upon $\gamma$ and $\delta$ as $m$ is kept constant. At $k_z=\pm\pi$, EPs annihilate each other to change their number. So, we solve the dispersion relation at $k_z=\pm\pi$ to obtain the phase space trajectory. Solving the dispersion relation of $H(k)_{NHW}$ at $k_z=\pm\pi$ , we get phase separation trajectory $(1+2\gamma\pm\delta)^2=2m^2-\delta^2$.  
 
We evaluate the value of $\delta$ from the phase space trajectory as,

\begin{equation}\label{eq2}
   \left\{\delta \to \pm \frac{1}{2} \left(2 \gamma \pm \sqrt{4 m^2-(2 \gamma +1)^2} + 1\right)\right\}.
\end{equation}

For a fixed value of $m$, $\delta$ depends only on $\gamma$, therefore, we obtain two intersecting ellipses in the $\gamma-\delta$ plane. This plot is shown in Fig. \ref{phase}, with the number of EPs in the bulk encoded with color code.  Notably, there exists a peculiar point in the parameter space $\gamma=-0.5$ and $\delta\sim0.4$ where an infinitesimal change in the value of $\delta$ or $\gamma$ will take to phases with different number of EPs therefore we coin it as triple point of EPs. 

\section{Surface states Analysis }
In this section, we calculate the tight-binding Hamiltonian of $H_\text{NHW} (k)$ with one axis OBC while maintaining PBC in the other two directions. The lattice version of the bulk Hamiltonian has the following form, 
\begin{equation}\label{seq4}
            H_{slab}=  H_0 \sum_{x=1}^N \, c_{x}^{\dagger}c_{x} +  \sum_{x=1}^{N-1} (t_x\,c_{x}^{\dagger}c_{x+1}+ \text{H.c.} ).
        \end{equation}
 Here, $H_0(k_z,k_y,\gamma,\delta,m)=(\gamma+\frac{1}{2}cos(k_z)\Gamma_4+  (\gamma +\frac{1}{2}cos(k_z)+ cos k_y)\Gamma_2 + sin k_y \Gamma_1 + m\sigma_0\kappa_2.$ and $t_x=\frac{1}{2}\Gamma_4-i\Gamma_3$ and $j$ is the unit cell index, we remind the readers that each unit cell has four spinless orbitals. The matrix form or the slab Hamiltonian of the above model is expressed in block off-diagonal form as,
\begin{equation}\label{seq5}
    H_{slab} = \begin{bmatrix} 
    H_0 & t_x & 0 & \dots & 0 & 0 & 0\\
    t_x^{\dagger} & H_0 & t_x & \dots & 0 & 0 & 0 \\
    0 & t_x^{\dagger} & H_0 & \dots & 0 & 0 & 0 \\
    \vdots & \vdots & \vdots & \ddots & \vdots &\vdots &\vdots \\
    0 & 0 & 0 & \dots & H_0 & t_x & 0 \\
    0 & 0 & 0 & \dots & t_x^{\dagger} & H_0 & t_x \\
    0 & 0 & 0 & \dots & 0 & t_x^{\dagger} & H_0 \\
    \end{bmatrix}.
\end{equation}
\begin{figure}[t]
    \centering
    \begin{tabular}{c c c c}
    \includegraphics[width=0.24\linewidth]{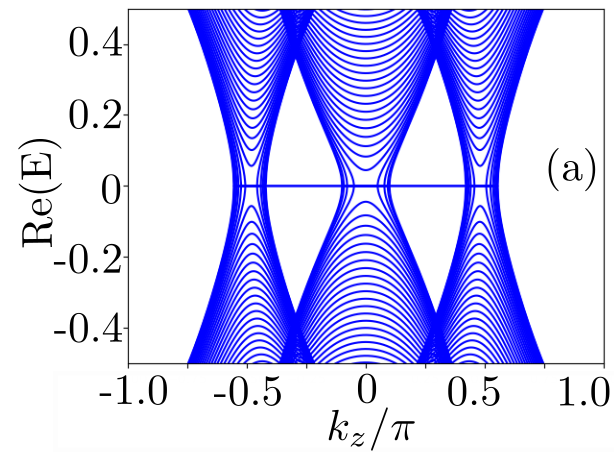}&
    \includegraphics[width=0.24\linewidth]{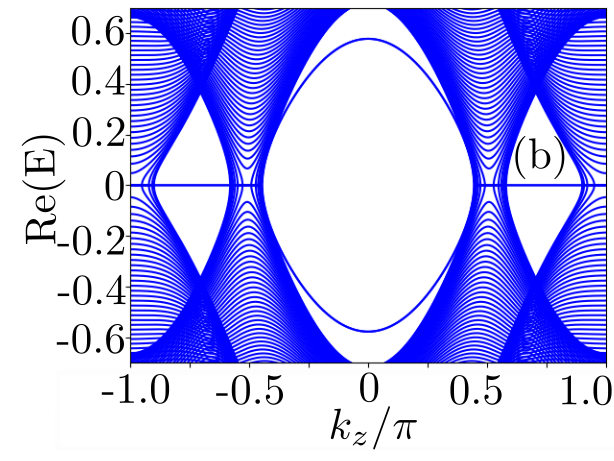}&
    \hspace*{-0.2cm}\includegraphics[width=0.24\linewidth]{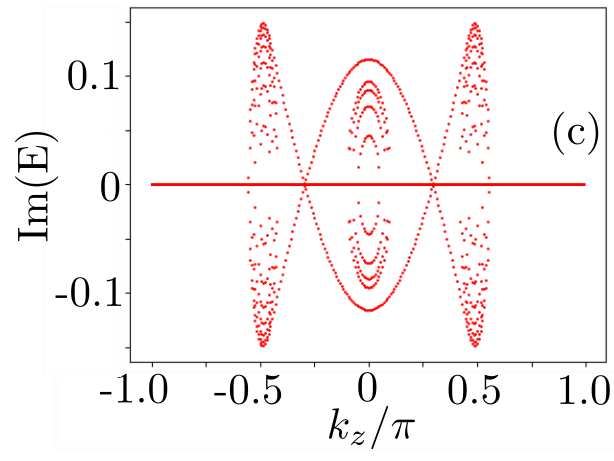}&
    \hspace*{-0.3cm}\includegraphics[width=0.24\linewidth]{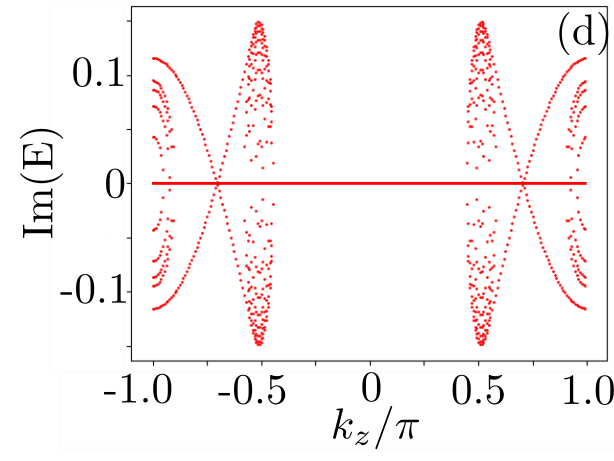}
    \\
    \includegraphics[width=0.24\linewidth]{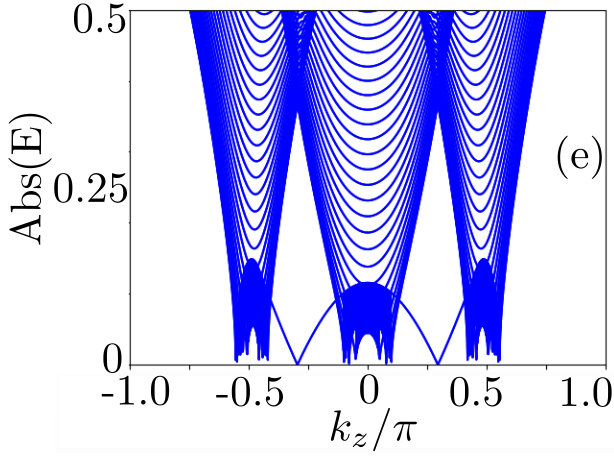}&
    \hspace*{-0.4cm}\includegraphics[width=0.24\linewidth]{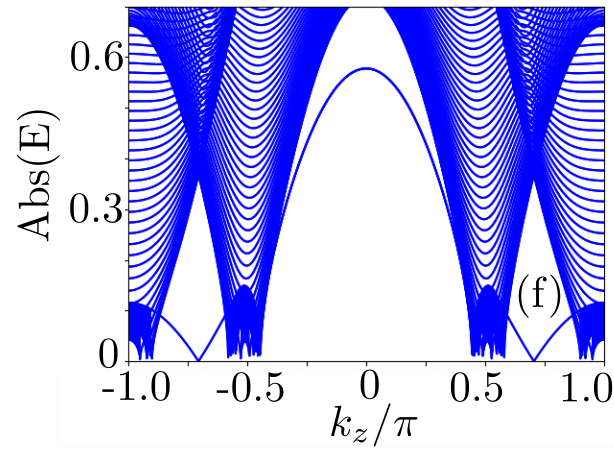}&
    \includegraphics[width=0.23\linewidth]{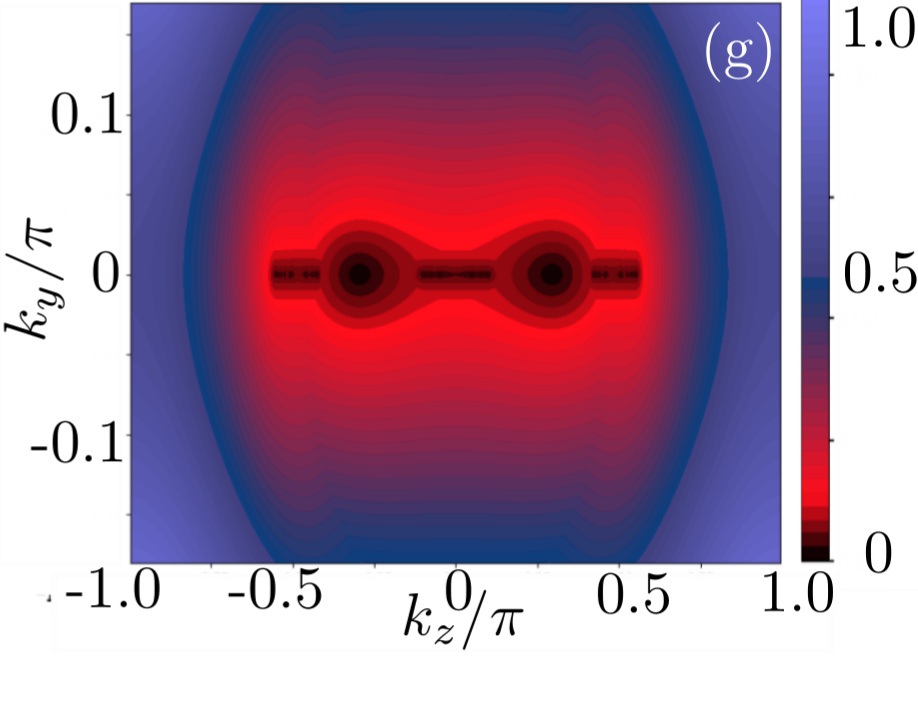} &
    \hspace*{-0.cm}\includegraphics[width=0.235\linewidth]{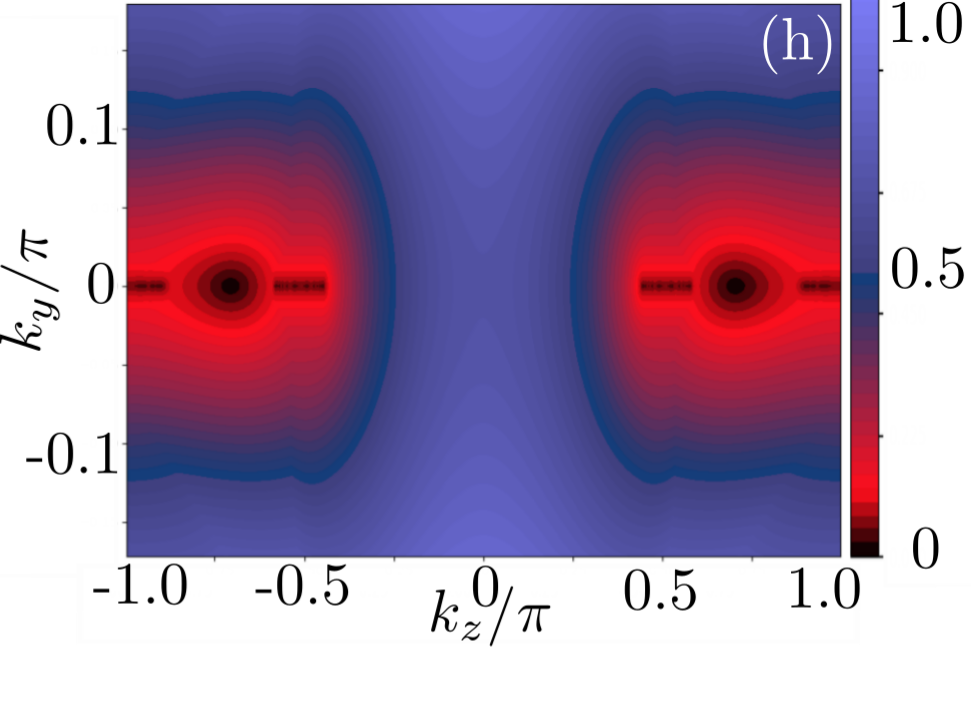}
    \end{tabular}
    \caption{Band diagrams for the slab Hamiltonian in Eq. (\ref{seq4}) as a function of  $k_z$. (a) The real part of the energy spectrum for $\gamma=-1.3$.   (b) Same plot for the parameter $\gamma=-0.7$. Both of them host zero energy surface states. (c) and (d) imaginary energy spectra for (a) and (b) respectively, both show the formation of concentric rings. 
    (e) and (f) Plots of the absolute value of energy for $\gamma=-1.3$ and $-0.7$ respectively, one can clearly point out the deviation of dispersion around DPs from  linear Dirac nature ($\gamma=-1.0$ in the main text). (g) and (h) are contour plots for (e) and (f) respectively, in ($k_y$-$k_z$) plane. These plots capture the nonlinear nature of the dispersion around the DPs (elliptical contour), which appear at $k_y=0 $ and  $k_z=\text{cos}^{-1}[{-2(1+\gamma)}]$.}
    \label{sur1}
\end{figure}
The diagonal or the onsite part of the slab Hamiltonian is,
\begin{equation}\label{seq6}
    H_0(k_z,k_y,\gamma,\delta,m)= 
    \normalsize{\left(
\begin{array}{cccc}
 0 & i \delta -i m & \gamma +\frac{\cos \left(k_z\right)}{2} & \gamma +e^{i k_y}+\frac{\cos \left(k_z\right)}{2} \\
 i \delta +i m & 0 & -\gamma -e^{-i k_y}-\frac{\cos \left(k_z\right)}{2} & \gamma +\frac{\cos \left(k_z\right)}{2} \\
 \gamma +\frac{\cos \left(k_z\right)}{2} & -\gamma -e^{i k_y}-\frac{\cos \left(k_z\right)}{2} & 0 & i \delta -i m \\
 \gamma +e^{-i k_y}+\frac{\cos \left(k_z\right)}{2} & \gamma +\frac{\cos \left(k_z\right)}{2} & i \delta +i m & 0 \\
\end{array}
\right)},
\end{equation}
whereas, the off-diagonal part is,
\begin{equation}\label{seq7}
    t_x= \left(
\begin{array}{cccc}
 0 & 0 & 1 & 0 \\
 0 & 0 & 0 & 0 \\
 0 & 0 & 0 & 0 \\
 0 & 1 & 0 & 0 \\
\end{array}
\right) \,\,\,\,\,\,\,\,\, and \,\,\,\,\,\,\,\,\,
t_x^{\dagger}= \left(
\begin{array}{cccc}
 0 & 0 & 0 & 0 \\
 0 & 0 & 0 & 1 \\
 1 & 0 & 0 & 0 \\
 0 & 0 & 0 & 0 \\
\end{array}
\right).
\end{equation}
The surface plots for the other two parameters, apart from $\gamma=-1$, are shown in Fig. \ref{sur1}.

\section{Higher order hinge state analysis }
\begin{figure}[t]
    \centering
    \begin{tabular}{c c c c}
    \includegraphics[width=0.23\linewidth]{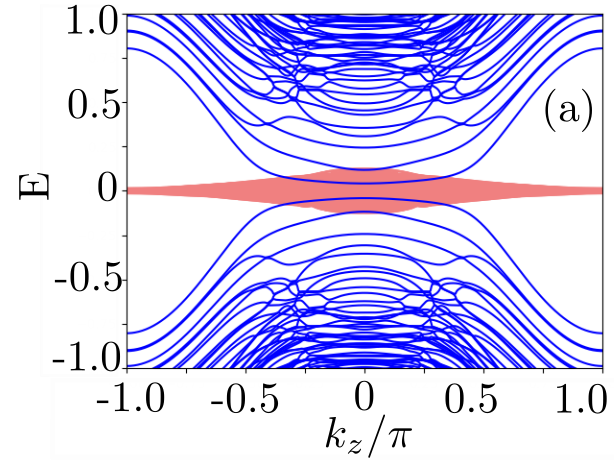}&
    \hspace*{-0.2cm}\includegraphics[width=0.23\linewidth]{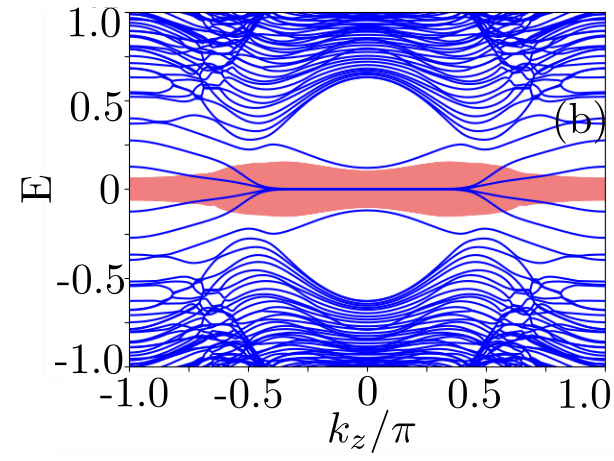}&
    \hspace*{-0.2cm}\includegraphics[width=0.23\linewidth]{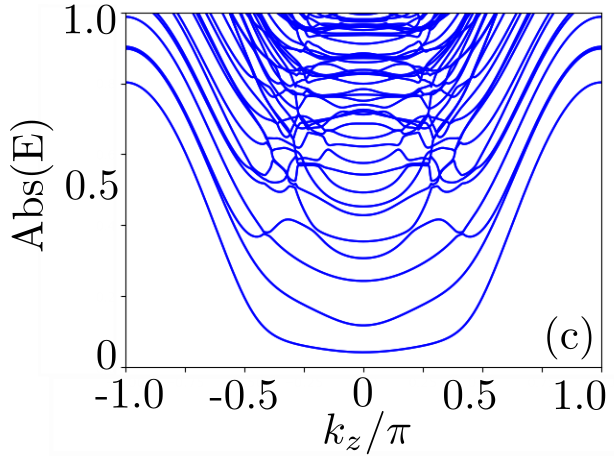}&
    \hspace*{-0.2cm}\includegraphics[width=0.23\linewidth]{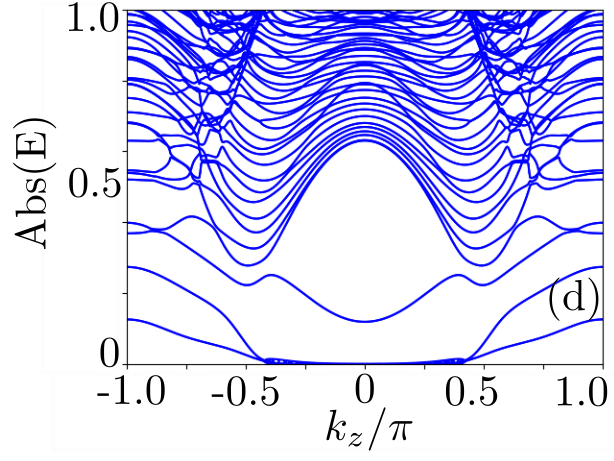}\\
    \includegraphics[width=0.22\linewidth]{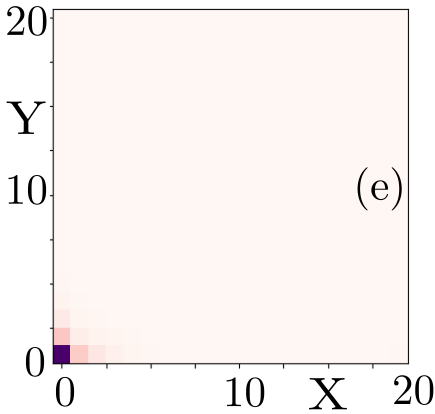}&
    \hspace*{-0.4cm}\includegraphics[width=0.225\linewidth]{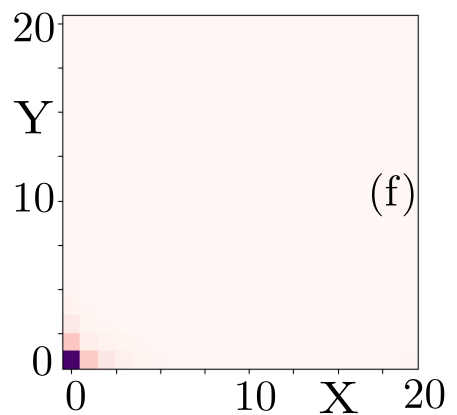}&
    \hspace*{-0.4cm}
    {\includegraphics[width=0.22\linewidth]{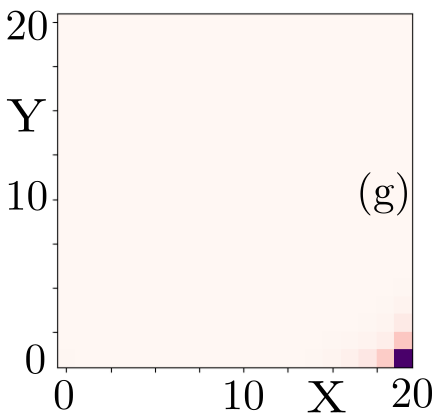}}&
    \hspace*{-0.4cm}\includegraphics[width=0.26\linewidth]{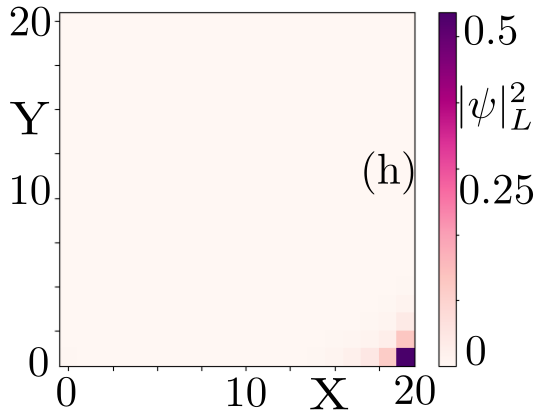}\\
    \includegraphics[width=0.22\linewidth]{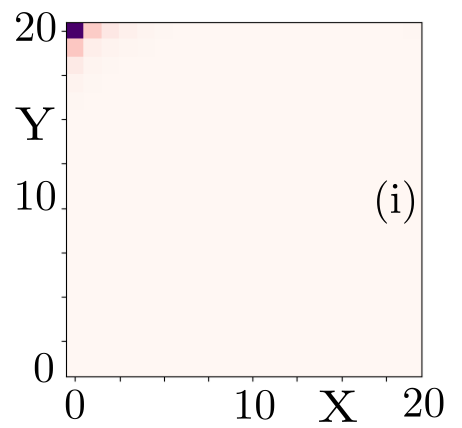}&
    \hspace*{-0.4cm}\includegraphics[width=0.215\linewidth]{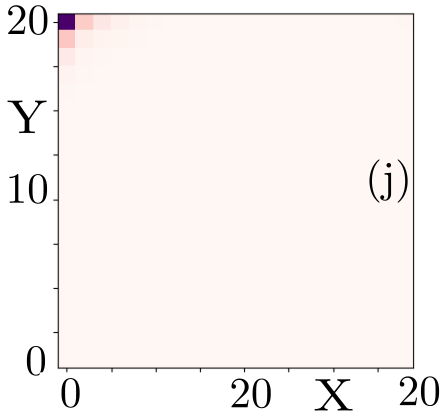}&
    \hspace*{-0.4cm}\includegraphics[width=0.22\linewidth]{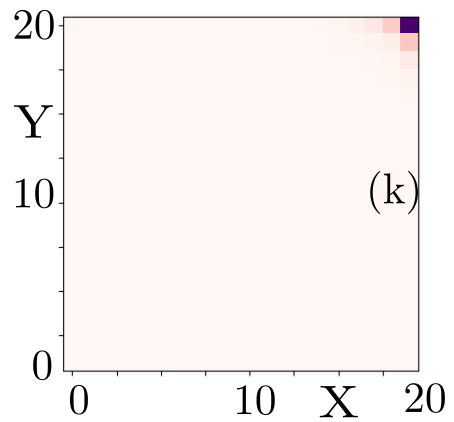}&
    \hspace*{-0.4cm}\includegraphics[width=0.26\linewidth]{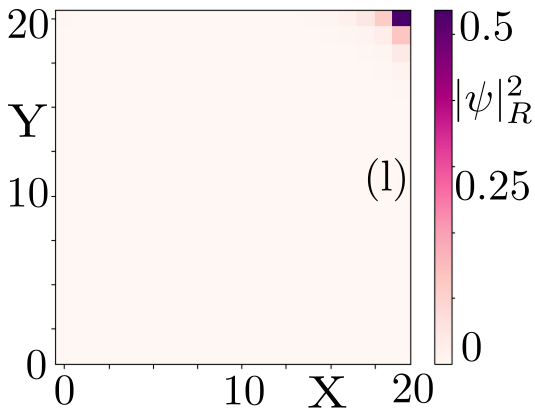}
    \end{tabular}
    \caption{The real (blue) and the imaginary (red) part of the energy eigen spectrum for two-axes OBC Hamiltonian for (a) $\gamma=-1.3$ and (b) $\gamma=-0.7$ as a function of $k_z$. In (b) hinge states appear connecting two EPs (higher-order) and therefore, for this parameter value $H_\text{NHW}$ is in the higher order topological phase (c) and (d) corresponding absolute energy plots for (a) and (b). (e)-(h) four hinge states left handed wavefunction $\ket{\psi_L}$ evaluated for $k_z=0$ for $\gamma=-0.7$ . Out of four, there are two distinct-looking wavefunctions forced to be caged at the lower two corners because of HONHSE. (i)-(l) The corresponding right wavefunction $\ket{\psi_R}$ evaluated in same region for the same parameter value. Remarkably, these are localized at the upper two corners. }
    \label{hinges}
\end{figure}
In order to construct the hinges of the bulk system, we apply OBC along the $x$ and $y$ axes however, PBC is maintained along the z direction. The lattice Hamiltonian in two directions takes the form, 
\begin{equation}\label{seq8}
            H_\text{hinge}= H_0 \sum_{x,y=1,1}^{N_x,N_y} \, c_{x,y}^{\dagger}c_{x,y} \, \ \delta_{x,y} + \Big( t_x \sum_{x,y=1,1}^{N_x-1,N_y} \,c_{x,y}^{\dagger}c_{x+1,y}+ t_y \sum_{x,y=1,1}^{N_x,N_y-1} c_{x,y}^{\dagger}c_{x,y+1}+ \text{H.c.} \Big),
\end{equation}
where $N_x=N_y=N$ is the number of unit cells along $x$ and $y$ directions on a square lattice. The form of $H_0$ has the following form
\begin{equation}\label{seq9}
    H_0(k_z,\gamma,\delta,m)= \left(
\begin{array}{cccc}
 0 & i \delta -i m & \gamma +\frac{\cos \left(k_z\right)}{2} & \gamma +\frac{\cos \left(k_z\right)}{2} \\
 i \delta +i m & 0 & -\gamma -\frac{\cos \left(k_z\right)}{2} & \gamma +\frac{\cos \left(k_z\right)}{2} \\
 \gamma +\frac{\cos \left(k_z\right)}{2} & -\gamma -\frac{\cos \left(k_z\right)}{2} & 0 & i \delta -i m \\
 \gamma +\frac{\cos \left(k_z\right)}{2} & \gamma +\frac{\cos \left(k_z\right)}{2} & i \delta +i m & 0 \\
\end{array}
\right),
\end{equation}
and the hopping matrices are,
\begin{equation}\label{seq10}
  t_x=\left(
\begin{array}{cccc}
 0 & 0 & 1 & 0 \\
 0 & 0 & 0 & 0 \\
 0 & 0 & 0 & 0 \\
 0 & 1 & 0 & 0 \\
\end{array}
\right) \,\,\,\,\,\,\,\,\,\text{and} \,\,\,\,\,\,\,\,\,t_y=\left(
\begin{array}{cccc}
 0 & 0 & 0 & 1 \\
 0 & 0 & 0 & 0 \\
 0 & -1 & 0 & 0 \\
 0 & 0 & 0 & 0 \\
\end{array}
\right).
\end{equation}

The band diagram figure for the Hamiltonian $H_{\text{hinge}}$ depicts that for $\gamma=-1.3$, the hinges are fully gapped however, for $\gamma=-0.7$, hinge states connect the two EPs closest to $k_z=0$. This marks the higher-order topological phase of $H_{NHW}$. We check that these hinge states are four-fold degenerate with absolute zero energy eigenvalues. The energy spectrum and eigenstates are shown in Fig. \ref{hinges}. Notably, the hinge states occupy only the lower corners. This mismatch lies in the fact that this model possesses  non-hermitian higher order skin effect (NHHOSE) \cite{zhang2021observation}  due to which hinges are caged in the lower corners only. In order to visualize all four hinge localizations, we plot the  left $\ket{\psi_L}$ along with  right $\ket{\psi_R}$ wavefunctions, as discussed in the main text.
\section{Slab Calculation for diabolic points: An analytical way to match eigenvalues}
In this section, we provide an analytical way of calculating eigenvalues corresponding to DPs and their degeneracy. We begin by using the slab Hamiltonian \cite{ETI} in Eq. (\ref{seq4}),  
\begin{equation}\label{seq11}
    H_{slab} = \begin{bmatrix} 
    H_0 & t_x & 0 & \dots & 0 & 0 & 0\\
    t_x^{\dagger} & H_0 & t_x & \dots & 0 & 0 & 0 \\
    0 & t_x^{\dagger} & H_0 & \dots & 0 & 0 & 0 \\
    \vdots & \vdots & \vdots & \ddots & \vdots &\vdots &\vdots \\
    0 & 0 & 0 & \dots & H_0 & t_x & 0 \\
    0 & 0 & 0 & \dots & t_x^{\dagger} & H_0 & t_x \\
    0 & 0 & 0 & \dots & 0 & t_x^{\dagger} & H_0 \\
    \end{bmatrix}.
\end{equation}
In the diagonal or onsite part of the above Hamiltonian, we make the following substitutions, $k_z=\pi/2$, $k_y=0$, $\gamma=-1$ and $m=2\delta$ (we set $\delta$ and m as used in the main text). After substituting and simplifying it looks like, 
\begin{equation}\label{seq12}
    H_0(k_z=\pi/2,k_y=0,\gamma=-1,\delta,m=2\delta)= 
    \left(
\begin{array}{cccc}
 0  & -i \delta  & -1 & 0 \\
 3 i \delta  & 0  & 0 & -1 \\
 -1 & 0 & 0  & -i \delta  \\
 0 & -1 & 3 i \delta  & 0  \\
\end{array}
\right).
\end{equation}
To derive the spectrum for these parameters, we analyze the solution of characteristics polynomial of $\text{det}[H_{\text{slab}}-\lambda \mathbb{I}_{4N \text{x} 4N}]=0$ where square lattice size is $4N$x$4N$. Even after these simplifications, the analytical solution remains infeasible. Therefore, to tackle this problem, we use Schur's determinant identity, \cite{terrier2020dissipative} 
\begin{equation}\label{seq13}
    det \left(
\begin{array}{cc}
 a_2 & t_x^{\dagger} \\
 t_x & a_1 \\
\end{array}
\right) = det(a_1) \, det(a_1-t_x a_1^{-1}t_x^{\dagger} ).
\end{equation}
\begin{figure}[t]
    \centering
    \includegraphics[width=8cm]{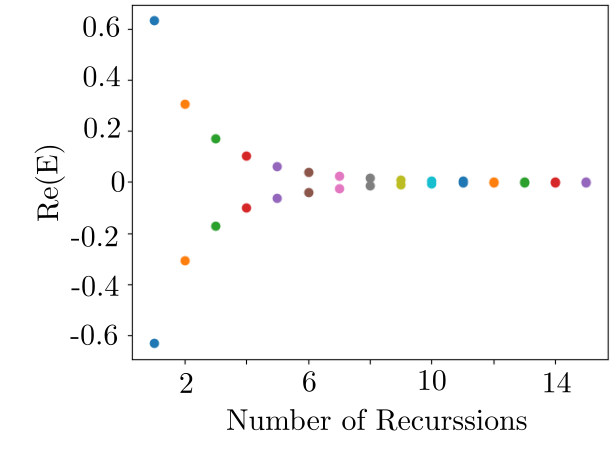}
    \caption{Two lowest energy eigenvalues as a function of the number of times Schur's determinant identity is used. We recursively use this identity and show that with the increase in the lattice size these two eigenvalues ultimately converge to zero.  This captures the degeneracy of the DPs as well as the finite size effect of the system.}
    \label{finite}
\end{figure}
Here, we set $a_1=H_0-\lambda \mathbb{I}_{4 X 4}$ and thus, for slab Hamiltonian $a_2=H_{slab}(\pi/2,0,-1,\delta,m=2\delta)_{4(N-1) \text{x} 4(N-1)}$. We  apply Schur's determinant identity to $a_2$ in an iterative manner and obtain  a recursion relation that allows us to calculate all eigenvalues. We can start with $\text{det}(a_1)=0$, which gives four solutions. Next we compute $a_2=a_1-(a_1-t_x a_1^{-1} t_x^{\dagger} )$ along with $\text{det}(a_2)=0$ and repeat this process recursively as, 
\begin{equation}\label{seq14}
    \boxed{
    a_{i+1}=a_1-(t_x a_{i}^{-1}t_x^{\dagger})}.
\end{equation}
\normalsize{
Finally, we  take $\text{det}(a_{i+1})=0$, which  gives us the eigenspectrum. 
We notice that there are two eigenvalues which converge to zero after continuing a series of iterations. The plot of the lowest energy solution with the number of times Schur's identity is being used is shown in Fig. \ref{finite}. Ultimately these two values converge to zero corresponding to two-fold degeneracy of DPs. It also captures the finite size effect, which demands the lattice to be large enough such that the DPs are exactly at zero energy.
\section{Derivation of momenta values at DPs}
In this section, we provide an analytical solution to find the exact momenta values at which the DPs are formed. 
In principle, one should work with the slab Hamiltonian, however, with $H_\text{slab}$ the analytical solution remains infeasible. Therefore, to proceed further, we project the bulk Hamiltonian along $k_x=0$ and $k_y=0$ momenta axes, 

\begin{equation}\label{seq15}
H(k_z,\gamma,\delta,m)=
    \left(
\begin{array}{cccc}
 0 & i \delta -i m & \gamma +\frac{\cos \left(k_z\right)}{2}+1 & \gamma +\frac{\cos \left(k_z\right)}{2}+1 \\
 i \delta +i m & 0 & -\gamma -\frac{\cos \left(k_z\right)}{2}-1 & \gamma +\frac{\cos \left(k_z\right)}{2}+1 \\
 \gamma +\frac{\cos \left(k_z\right)}{2}+1 & -\gamma -\frac{\cos \left(k_z\right)}{2}-1 & 0 & i \delta -i m \\
 \gamma +\frac{\cos \left(k_z\right)}{2}+1 & \gamma +\frac{\cos \left(k_z\right)}{2}+1 & i \delta +i m & 0 \\
\end{array}
\right).
\end{equation}
\normalsize
To reduce one more variable we keep $m=2\delta$ as in the main text. The characteristics polynomial of the above projected Hamiltonian is given by the form, 

\begin{equation}\label{seq16}
    E^4  +  A \,E^2  +B =0,
\end{equation}
\normalsize
where E is the energy eigenvalue, A and B have the following forms
\begin{equation}
    A \,\,=\,\, 
    \left(\delta ^2-2 \gamma  (\gamma +2)\right)-8 \delta ^2-\cos \left(k_z\right) \left(4 \gamma +\cos
   \left(k_z\right)+4\right)-4,
\end{equation}  and

\begin{equation}
\begin{split}
    B\,\,=\,\,-16 \gamma  (\gamma +2) \delta ^2+4 \gamma  (\gamma +2) (\gamma  (\gamma +2)+2)+9 \delta ^4-16 \delta ^2-4 \delta ^2
   \cos \left(k_z\right) \left(4 \gamma +\cos \left(k_z\right)+4\right)  \\
   +\frac{1}{4} \cos ^2\left(k_z\right) \left(4
   \gamma +\cos \left(k_z\right)+4\right){}^2+2 (\gamma +1)^2 \cos \left(k_z\right) \left(4 \gamma +\cos
   \left(k_z\right)+4\right)+4.
\end{split}
\end{equation}
\normalsize
We solve the characteristics polynomial for low energy, thus  putting $E^4=0$ and solving for $E^2=0$, which is satisfied at $B/A=0$. After simplification we get,
\begin{equation}
    B/A= -\gamma  (\gamma +2)+\frac{11 \delta ^2}{2}-\frac{42 \delta ^4}{4 (\gamma +1)^2+6 \delta ^2+\cos \left(k_z\right)
   \left(4 \gamma +\cos \left(k_z\right)+4\right)}-\frac{1}{4} \cos \left(k_z\right) \left(4 \gamma +\cos
   \left(k_z\right)+4\right)-1.
\end{equation}
Neglecting the second and third terms as $\delta$ is small enough, it further simplifies into, 
\begin{equation}
    B/A= -\gamma  (\gamma +2)-\frac{1}{4} \cos \left(k_z\right) \left(4 \gamma +\cos \left(k_z\right)+4\right)-1.
\end{equation}
Solving  $B/A=0$ leads us to the following values of  $k_z$

\begin{equation}
   \boxed{k_z\to \pm \cos ^{-1}[-2 (\gamma +1)]}.
\end{equation}

We note that $k_z$ is independent of $m$ and $\delta$ and explicitly depends only on $\gamma$. Also, the DPs appear exactly at those locations  where the Dirac nodes are formed in the HODS \cite{huges} phase. From this, one can clearly deduce that DPs retain the memory of HODS phase.\\
\section{Verification of low energy dispersion relation near diabolic points}
 The Hamiltonian $H_\text{NHW}$ in the main text has the following matrix form, 
 \begin{equation}
    H_\text{NHW}(k_x,k_y,k_z,\gamma,\delta,m)= 
    {\left(
\begin{array}{cccc}
 0 & i \delta -i m & \gamma +e^{i k_x}+\frac{\cos \left(k_z\right)}{2} & \gamma +e^{i k_y}+\frac{\cos \left(k_z\right)}{2} \\
 i \delta +i m & 0 & -\gamma -e^{-i k_y}-\frac{\cos \left(k_z\right)}{2} & \gamma+e^{-i k_x} +\frac{\cos \left(k_z\right)}{2} \\
 \gamma +e^{-i k_x}+\frac{\cos \left(k_z\right)}{2} & -\gamma -e^{i k_y}-\frac{\cos \left(k_z\right)}{2} & 0 & i \delta -i m \\
 \gamma +e^{-i k_y}+\frac{\cos \left(k_z\right)}{2} & \gamma +e^{i k_x} +\frac{\cos \left(k_z\right)}{2} & i \delta +i m & 0 \\
\end{array}
\right)}.
\label{S21}
\end{equation}
\normalsize{
We notice that the DPs are formed on the $x$-surface of the system. Therefore, we project the bulk Hamiltonian in the $k_y-k_z$ plane, assuming $k_x=0$.  In principle, one should work with the slab Hamiltonian, however, with $H_\text{slab}$ the analytical solution remains infeasible.  Therefore, we work only with $H_0$ (the diagonal part which depends on momentas) to extract the dependency of $E$ on $k_y$ and $k_z$.
First, we expand the Hamiltonian near these momenta values of DPs for small values of $k_y$ which gives $e^{\pm i k_y}=1\pm i k_y$. Second,  expansion of $k_z$ around $\pm \pi/2$ gives ${\cos \left(k_z\right)=\cos \left(k_{z}\pm \pi/2\right)=\pm \sin{(k_z)}=k_z}$. This step tremendously simplifies the Hamiltonian, thus the matrix form of $H_0$ takes the form}
\begin{equation}
    H_0(k_x=0,k_z,k_y,\gamma=-1,\delta,m=\delta/2)= 
    \normalsize{\left(
\begin{array}{cccc}
 0 &  -i\delta  & \frac{ k_z}{2} & i k_y+\frac{ k_z}{2} \\
 3i \delta  & 0 & i k_y-\frac{ k_z}{2} &  \frac{ k_z}{2} \\
 \frac{ k_z}{2} & -i k_y-\frac{ k_z}{2} & 0 & -i \delta\\
 -i k_y+\frac{ k_z}{2} & \frac{ \left(k_z\right)}{2} & 3i\delta  & 0 \\
\end{array}
\right)}.
\label{S22}
\end{equation}

\begin{figure}[h]
    \centering
    \includegraphics[width=6cm]{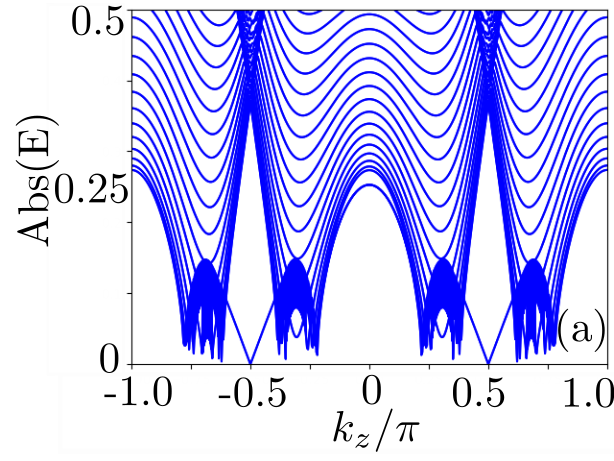}
    \includegraphics[width=6cm]{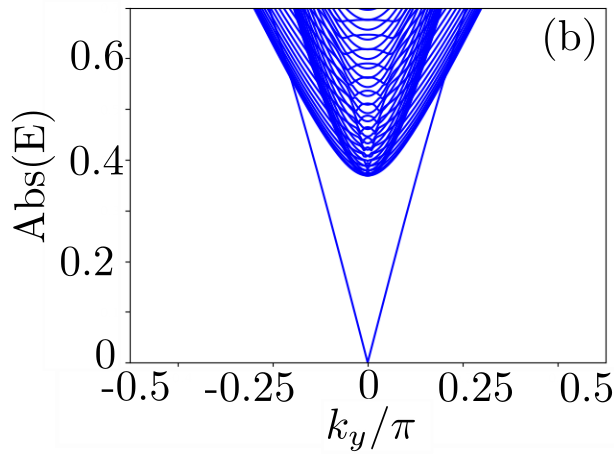}
    \caption{Absolute energy band diagram for the OBC (surface) Hamiltonian of the main text, (a) along $k_z$ ($k_y=0$) showing the formation of two Dirac cones with linear dispersion at low energy. These are formed exactly at $k_z=\pm \pi/2$. (b) surface band diagram with same parameters along $k_y$ ($k_z=\pm \pi/2$), depicting a single Dirac cone at $k_y=0$. At this particular parameter, that is, $\gamma=-1$ low energy dispersion relations are linear along both the momenta direction.}
    \label{dirac}
\end{figure}
Although this form of Hamiltonian looks simple, the analytical form of the eigenvalues is complicated. In order to reduce the complicated form of eigenvalues, we take $k_y=0$ and solve the dispersion relation for $k_z$ and vice-versa. First taking $k_y=0$, the dispersion relation along $k_z$ has the form
\begin{equation}
    E^4(k_z)= (k_z^2+2\sqrt{7} \delta k_z+6\, \delta^2) (k_z^2-2\sqrt{7}\delta k_z+6\, \delta^2)/4
    = \big[ (k_z^2+6\, \delta^2)^2 -28 \delta^2 k_z^2\big]/4
\end{equation}
and for $k_z=0$, along $k_y$ the dispersion takes the form, 
\begin{equation}
    E^4(k_y)=(k_y -3 i\,\delta) (k_y+i\,\delta)(k_y +3i\,\delta) (k_y-i\,\delta)=(k_y^2 +9\delta^2) (k_y^2+\delta^2).
\end{equation}
For small $\delta$, along both axes dispersion of the absolute value of energy $\sim |k|$ is shown in Fig. \ref{dirac}. Therefore, we verify analytically the linear form of low energy dispersion along $k_y$ and $k_z$ direction around the DPs at $\gamma=-1$.
 
On plotting the surface band diagrams, we notice that at $\gamma=-1.5$ or -0.5, the low energy dispersion around DPs is quadratic in nature along $k_z$ and linear along $k_y$. This kind of quadratic dispersion is seen in Luttinger semimetals \cite{mandal2021transport,boettcher2016superconducting,ghorashi2018irradiated}. To analytically capture the quadratic dependency, we make following substitution in Eq. (\ref{S21}), $k_x=0$, and expand close to $k_y=0$ and $k_z=0$ such that $e^{\pm i k_y}=1\pm i k_y$ and $\cos(k_z)=1+\frac{k_z^2}{2}$ after neglecting higher order terms. We note that  the only difference between $\gamma=-1.5$ and -0.5 is the value of $k_{z0}$. Further, some algebraic manipulations simplify the Hamiltonian into,}
\begin{equation}
    H_0(k_x=0,k_y,k_z,\gamma=-1.5,\delta,m=\delta/2)= 
    {\left(
\begin{array}{cccc}
 0 &  -i \delta  & \frac{ k_z^2}{4} & i k_y+\frac{ k_z^2}{4} \\
 3i \delta  & 0 & i k_y-\frac{ k_z^2}{4} &  \frac{ k_z^2}{4} \\
 \frac{ k_z^2}{4} & -i k_y-\frac{ k_z^2}{4} & 0 & -i \delta\\
 -i k_y+\frac{ k_z^2}{4} & \frac{ k_z^2}{4} & 3i \delta & 0 \\
\end{array}
\right)}.
\end{equation}
\begin{figure}[t]
    \centering
    \includegraphics[width=6cm]{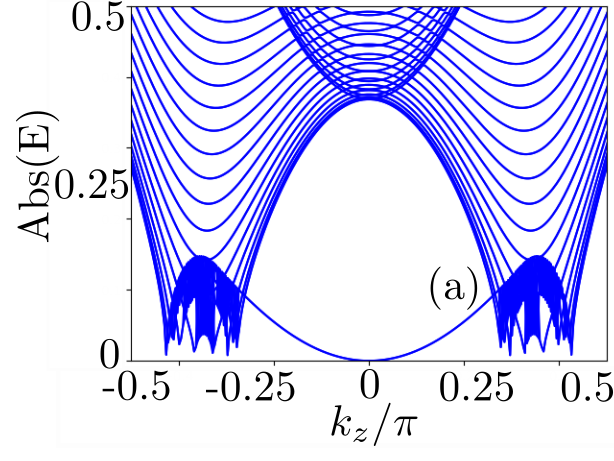}
    \includegraphics[width=6cm]{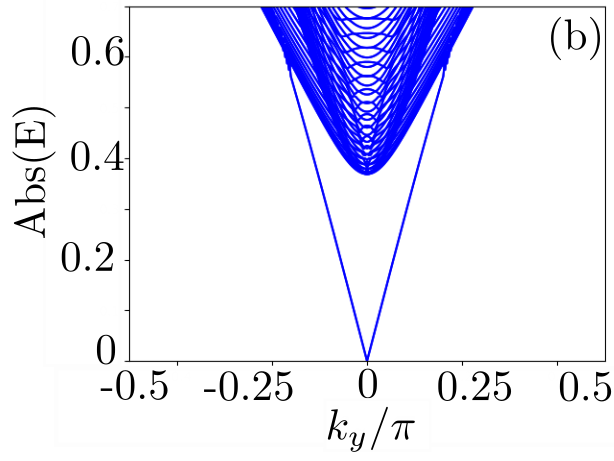}
    \caption{Surface band diagram similar to  Fig. \ref{dirac} with  $\gamma=-1.5$. (a) as a function of $k_z$ ($k_y=0$) depicting a Luttinger quadratic dispersion relation. At $\gamma=-0.5$ this  quadratic dispersion is formed at $k_z=\pm \pi/2$ as shown in Fig. 2(e) of the main text. (b) In the band diagram, a single Dirac cone has been observed as a function of $k_y$. Therefore, along $k_y$ the dispersion is always linear in nature.}
    \label{lutt}
\end{figure}
It has an exactly similar form to Eq. (\ref{S22}) where only $k_z/2$ gets replaced with $k_z^2/4$ and remarkably $ik_y$ remains the same.
 Therefore, using a similar analysis we can account for quadratic dispersion, with absolute value of $E \sim k_z^2$, along one axis and as $k_y$ along another. This kind of dispersion relation is captured in Fig. \ref{lutt}. 

\section{Topological Invariants}
\subsection{1-D winding number to characterize the topology of Bulk EPs}
The main text Hamiltonian $H_{NHW}(k)=H_0(k)+i\delta\sigma_0\kappa_1$, is not off-block diagonal for the chosen basis. So, in order to invoke the hidden sublattice symmetry in the model and make it an off-block diagonal, we change the basis of the Hamiltonian $H_{NHW}$ by rotating the Pauli matrix ($\sigma_s^i$ and $\kappa_s^i$) about the $y$ axis. The transformations are as follows, $\sigma^x \rightarrow \kappa^y\sigma^x$, $\sigma^y \rightarrow \sigma^y$, $\sigma^z \rightarrow \kappa^y\sigma^z$ for $\sigma_s^i$ and  $\kappa^x \rightarrow \sigma^y\kappa^x$, $\kappa^y \rightarrow \kappa^y$, $\kappa^z \rightarrow \sigma^y\kappa^z$ for $\kappa_s^i$. After successfully applying these transformations to the Hamiltonian and projecting it along the $x$-axis, the bulk Hamiltonian takes the form, 

\begin{equation}
    H_{NHW}(k)= \left(
\begin{array}{cc}
 0 & Q_1 \\
 Q_2 & 0 \\
\end{array}
\right)
\end{equation}
where the upper block is given by,
\begin{equation}
    Q_1=\left(
\begin{array}{cc}
 -\sin \left(k_y\right)-i \text{m} & \delta +(1-i) \left(\gamma +\frac{\cos \left(k_z\right)}{2}\right)-i+\cos \left(k_y\right) \\
 -\delta +(-i-1) \left(\gamma +\frac{\cos \left(k_z\right)}{2}\right)-i-\cos \left(k_y\right) & -\sin \left(k_y\right)-i \text{m} \\
\end{array}
\right),
\end{equation}
and, the lower block is given by,
\begin{equation}
    Q_2=\left(
\begin{array}{cc}
 -\sin \left(k_y\right)+i \text{m} & \delta +(i-1) \left(\gamma +\frac{\cos \left(k_z\right)}{2}\right)+i-\cos \left(k_y\right) \\
 -\delta +(i+1) \left(\gamma +\frac{\cos \left(k_z\right)}{2}\right)+i+\cos \left(k_y\right) & -\sin \left(k_y\right)+i \text{m} \\
\end{array}
\right).
\end{equation}
As the system is NH,  $Q_2 \neq Q_1^{\dagger}$ as expected. The winding number as a function of $k_z$ is given by,
\begin{equation}
     W_{(1,2)}^{k_x=0}(k_z)=\int_{-\pi}^{\pi}\frac{1}{2\pi i}\partial_{k_y}\log\big(\text{det}[ Q_{1,2}(k_y,k_z)]\big)dk_y.
\end{equation}
As discussed in the main text, it is noticed that $W=W_1+W_2$ flip sign after hitting any of the EPs and jumps from 0 to $\pm 1$. Therefore, we conclude that these bulk EPs are topological in nature.
\subsection{Real space open boundary Chern number }
In Hermitian systems, the appearance of robust topologically protected edge modes can be directly predicted only by looking at the bulk. These edge modes appear whenever there is a quantization of the Chern number or any of the topological indices in the bulk. So, if one knows about the bulk, the boundary can be predicted and vice versa, this is known as bulk boundary correspondence (BBC). This is a remarkable phenomenon in Hermitian systems where the topological phase transition is marked by band touchings. On the contrary, NH systems do not follow the traditional BBC \cite{genbbc,okuma2020topological}. In such a scenario, it is quite difficult to reestablish a simplified correspondence between  the bulk or boundary as in the case of Hermitian systems. Therefore, we calculate the topological invariant in real space with open boundaries as opposed to calculating them in momentum space. 
To begin with,  we use the eigenstates of the Hamiltonian with OBC along two axes. 
As written in the main text, using $\hat{P_{\alpha}}$, open boundary Chern number is calculated as \cite{song2019non},
\begin{equation}
    C_\alpha=\frac{2\pi \iota}{L_{x}'L_{y}'}Tr'\left(\hat{P_\alpha}[[\hat{X},\hat{P_\alpha}],[\hat{Y},\hat{P_\alpha}]]\right).
\end{equation}
\subsection{Biorthogonal higher order topological invariant: $q_{xy}$}
It is well known that the appearance of hinge states is a clear signature of the higher order topological phase of 3D systems\cite{huges, wang}. Hinge gapless modes appear where the one-axis OBC spectrum, that is, the surface is gapped. We notice that the innermost EPs closest to the  $\Gamma$ point in BZ are connected by four-fold degenerate hinge states. Thus, we conclude that the innermost EPs are higher-order \cite{mandal2021symmetry,hodaei2017enhanced}. For Hermitian systems as well as their NH generalizations, higher order topology is always characterized by the presence of quadruple moment, $q_{xy}$.  
This kind of analysis is well explored in existing literature \cite{multipole,benalelectric} and is given by, 
\begin{equation}
    Q_{corner}-p_x^{edge}-p_y^{edge}=-q_{xy}.
\end{equation}
Here, $p_x^{edge}$ ($p_y^{edge}$) is surface polarization along the $x (y)$ axis. The charge $Q_{corner}$  
 is localized at the corners of each 2D $k_z$ slice of the material. As mentioned in the previous section due to the presence of the NH skin effect and breaking down of BBC, it is challenging to use this formula \cite{nhhigherdirac} as  the explicit calculations of $p_x^{edge}$ and $p_y^{edge}$ and corner charge($Q_{corner}$) are complicated. In order to avoid all these complications and be motivated by the previous success of biorthogonal approach, we use real space operator-based formalism to calculate the quadruple moment $q_{xy}$.
From a previous study \cite{wu2021floquet}, using the same line of thought $q_{xy}$ is defined in main text as, 
\begin{equation}
     q_{xy}=\left (\frac{Im (\,\, \text{ln} \, [\text{det}s \,\, \hat{Q}] \, )}{2\pi}-\frac{\sum_i \hat{X}_i\hat{Y}_i}{2 l_x l_y}\right) \,\, mod \,\, 1,
\end{equation}

This formalism gives the quantized value of quadruple moment  to half only at those locations where gapless hinge modes appear connecting the innermost EPs.
\section{Model preserving Inversion symmetry}
\begin{figure}[h]
    \centering
\begin{tabular}{c c c }
     \includegraphics[width=0.28\linewidth]{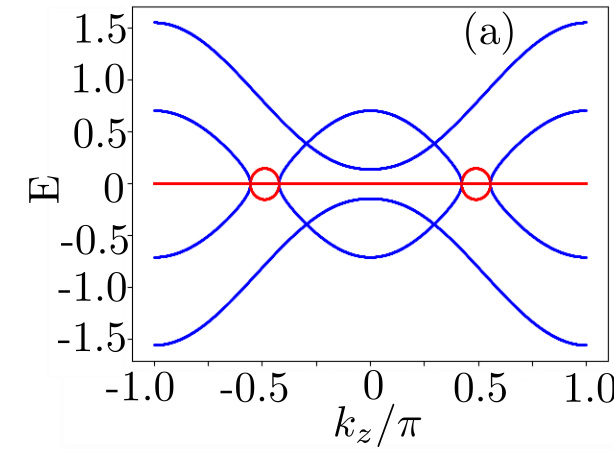}&
     \hspace*{-.1cm}\includegraphics[width=0.28\linewidth]{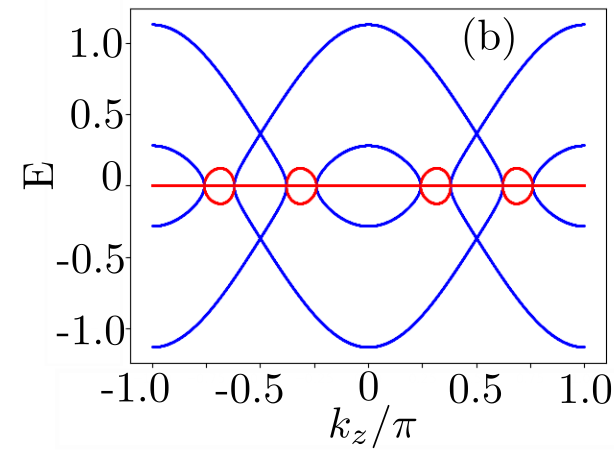}&
     \hspace*{-.1cm}\includegraphics[width=0.28\linewidth]{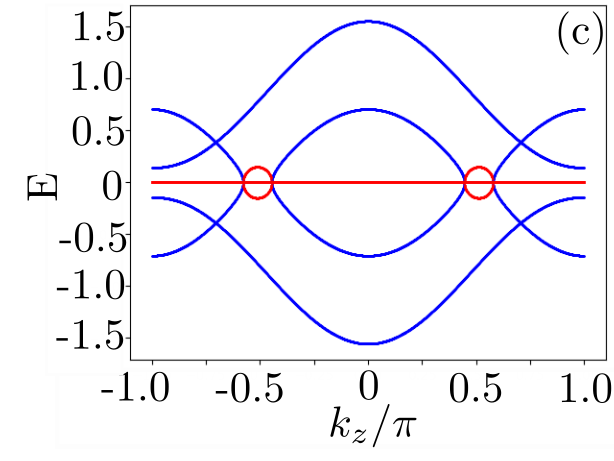} \\ 
     \includegraphics[width=0.28\linewidth]{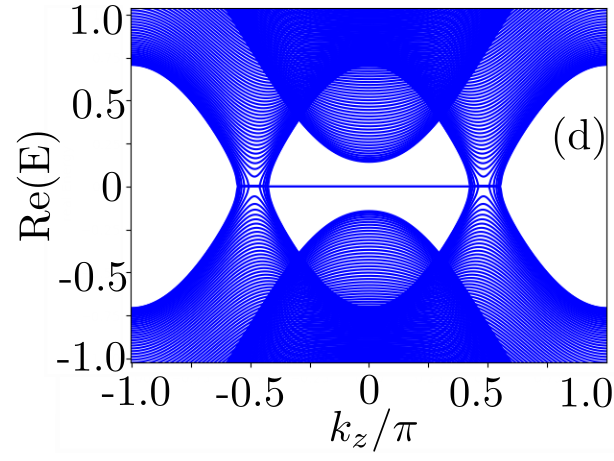} &
    \hspace*{-0.1cm} \includegraphics[width=0.28\linewidth]{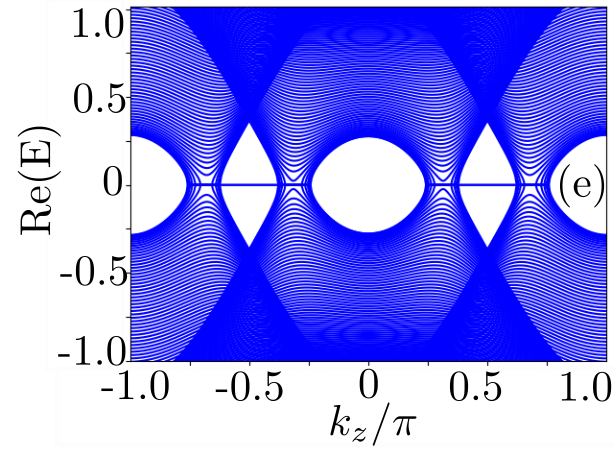}&
    \hspace*{-0.1cm}\includegraphics[width=0.28\linewidth]{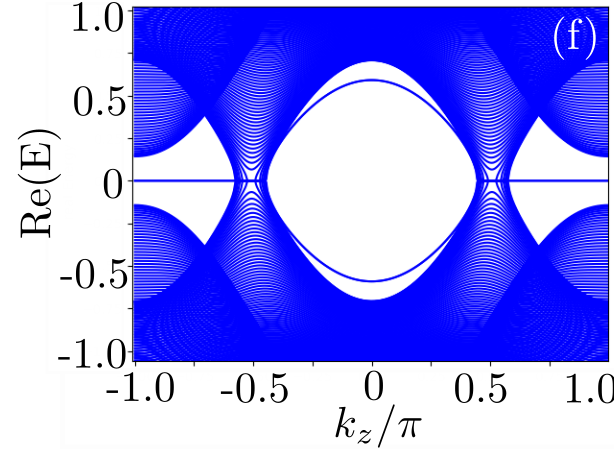} \\ 
    \includegraphics[width=0.28\linewidth]{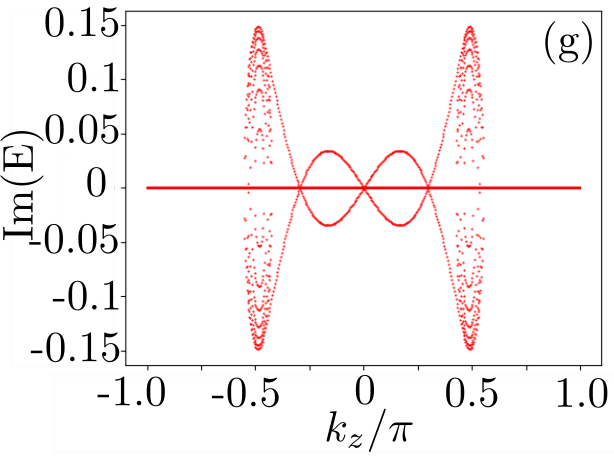}&
    \hspace*{-0.1cm}\includegraphics[width=0.28\linewidth]{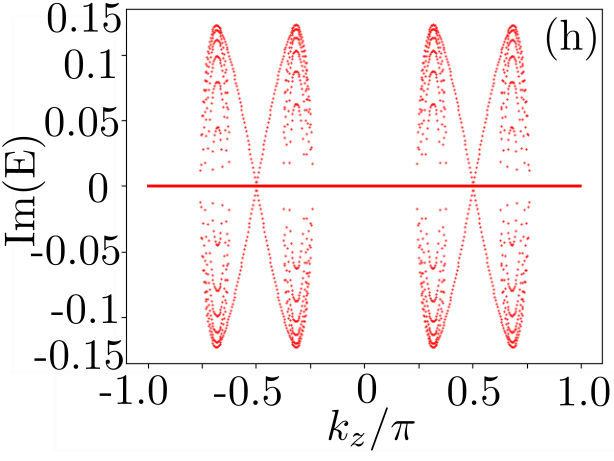} &
     \hspace*{-0.1cm}\includegraphics[width=0.28\linewidth]{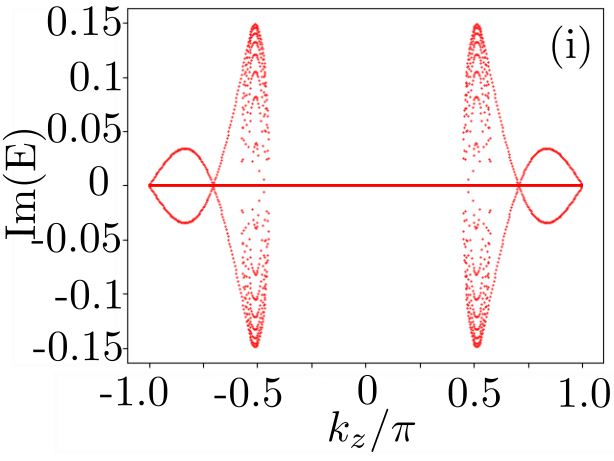} \\ 
     \includegraphics[width=0.28\linewidth]{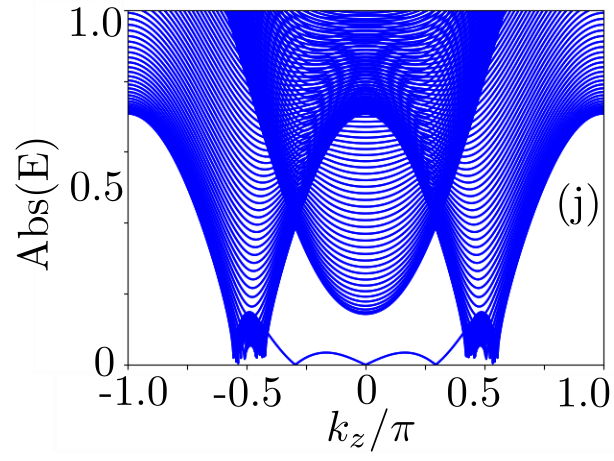}&
     \hspace*{-0.1cm}\includegraphics[width=0.28\linewidth]{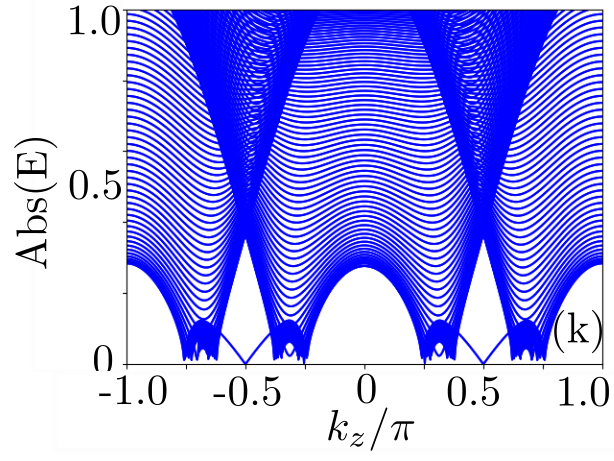}&
     \hspace*{-0.1cm}\includegraphics[width=0.28\linewidth]{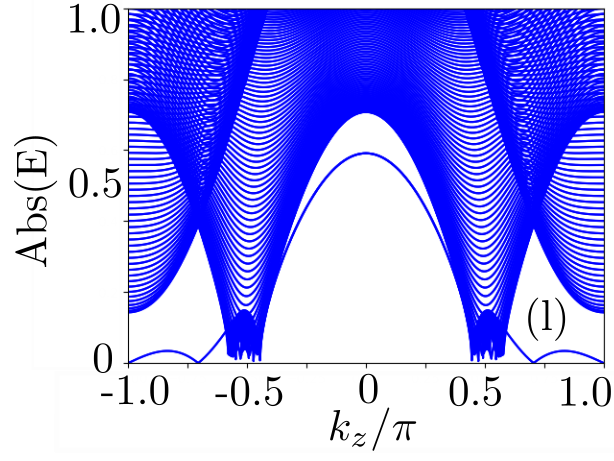} \\
          \includegraphics[width=0.28\linewidth]{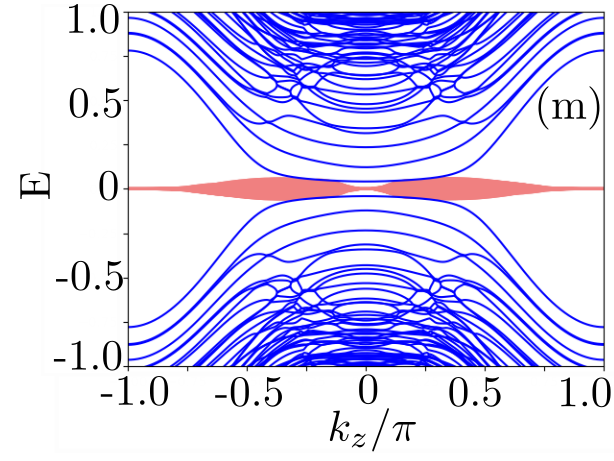} & 
          \hspace*{-0.1cm}\includegraphics[width=0.28\linewidth]{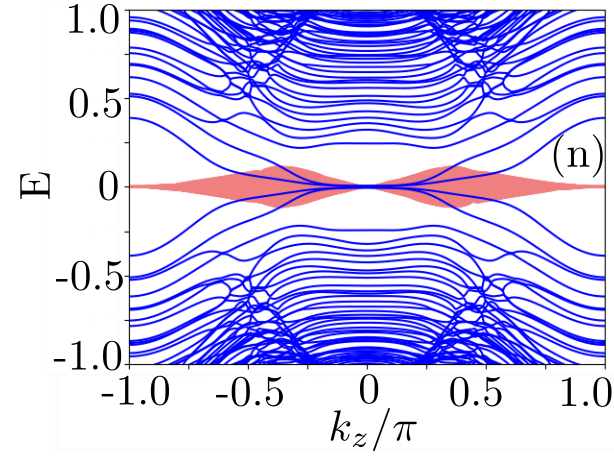}&
          \hspace*{-0.1cm}\includegraphics[width=0.28\linewidth]{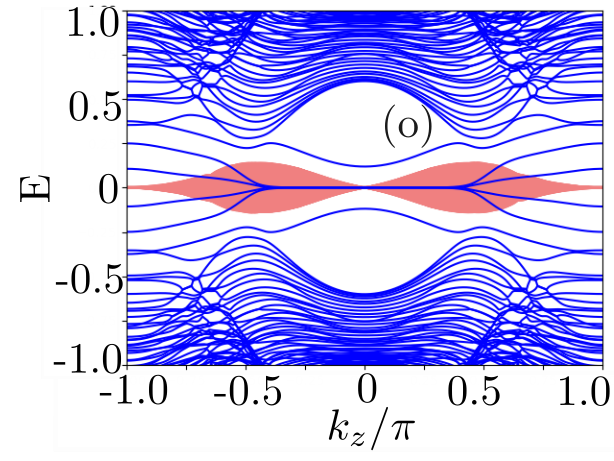} \\ \includegraphics[width=0.28\linewidth]{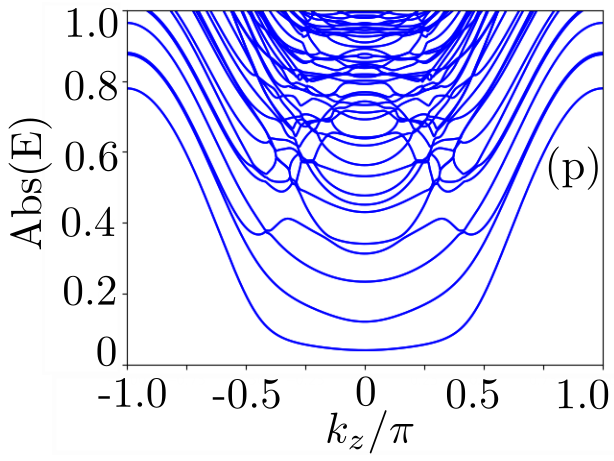} &
  \hspace*{-0.1cm}\includegraphics[width=0.28\linewidth]{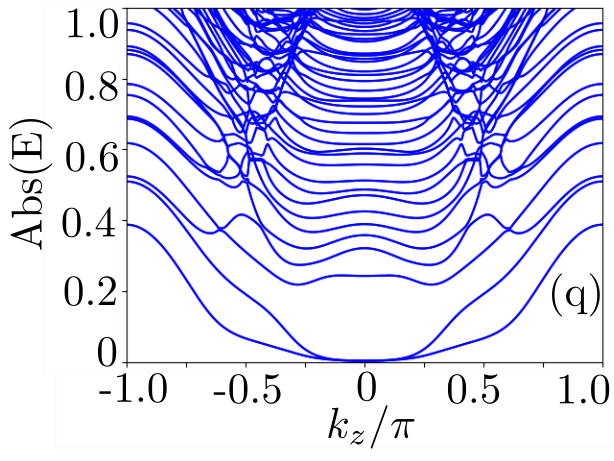}&
  \hspace*{-0.1cm}\includegraphics[width=0.28\linewidth]{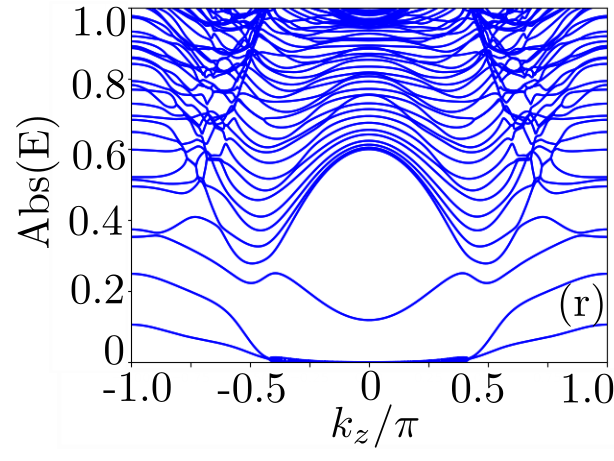}
\end{tabular}
   
    \caption{(a)-(c) Bulk energy spectrum for inversion symmetric Hamiltonian written in Eq. (\ref{IH}), for  $\gamma(-1.3,-1.0,-0.7)$ from left to right column in all the rows. We consider $m$ and $\delta$ to have the same values as in the main text.  Corresponding  real, imaginary,  and absolute parts of eigenvalues for the $x$ surface, have been shown in (d)-(f), (g) - (i), and (j)- (l), respectively. The absolute energy consists of DPs with modulating dispersion with the variation in $\gamma$. In (m)-(o), hinge spectrum with two-axes OBC for inversion symmetric model with real part in blue and imaginary in red. (p)-(r) Corresponding absolute energy spectrum for hinge states. Notably, spectrum is linear close to DPs at $\gamma=-1$.}
    \label{invsym}
\end{figure}

Here we present a model with explicit momentum dependence
in the NH perturbation. As a result, it breaks the time-reversal symmetry
($\mathcal{T}$) but preserves inversion symmetry ($\mathcal{P}$) \cite{hughesinv}.
\begin{equation}
    H^{\mathcal{I}}_{\text{NHW}}(k)=H_0(k)+i\, \delta\sigma_0\kappa_1 \sin(k_z),
    \label{IH}
\end{equation}
where $H_0(k)$ is the same Hermitian part as of the main text,
\begin{align}
    &H_0(k)=  \sum_{j=1}^4 h_j \Gamma_j +  m \sigma_0\kappa_2.
\end{align}
Similar to the model in the main text, here also we can control the number of EPs in the bulk. The EPs are either connected by first-order surface or higher-order hinge states, depending upon the topological phase of the material.  All the band diagrams corresponding to the bulk, surface, and hinges are plotted in Fig. \ref{invsym}. Upon increasing the value of $\gamma$, from -1.3 to -0.7, we are able to generate higher-order EPs in the bulk, which are connected by hinge states. Thus, similar to the model in the main text, one can traverse from normal to a higher-order phase by tuning system parameters.  

\section{Model preserving Time Reversal symmetry}
\begin{figure}[h]
    \centering
\begin{tabular}{c c c }
     \includegraphics[width=0.28\linewidth]{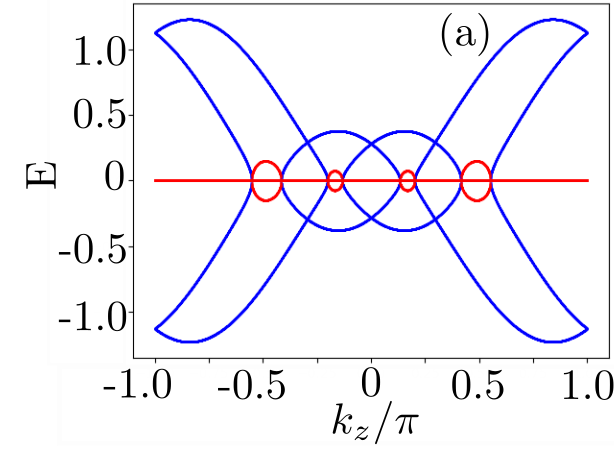}&
     \hspace*{-.1cm}\includegraphics[width=0.28\linewidth]{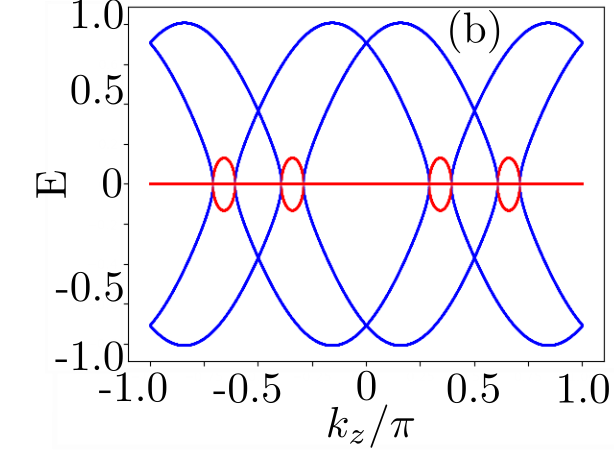}&
     \hspace*{-.1cm}\includegraphics[width=0.28\linewidth]{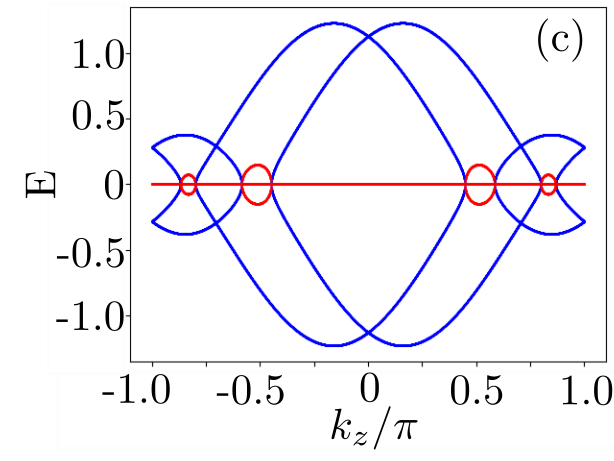} \\
     \includegraphics[width=0.28\linewidth]{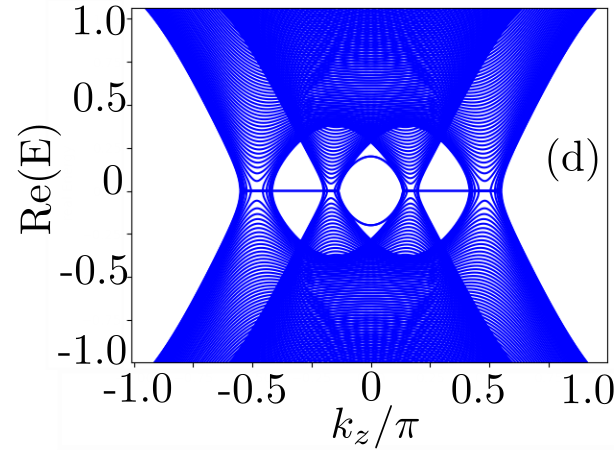} &
    \hspace*{-.1cm} \includegraphics[width=0.28\linewidth]{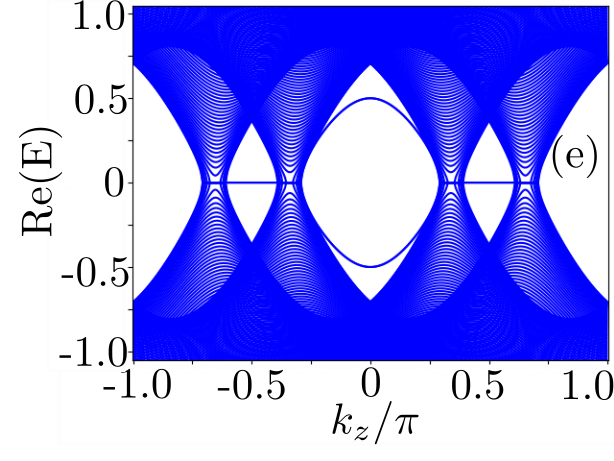}&
    \hspace*{-.1cm}\includegraphics[width=0.28\linewidth]{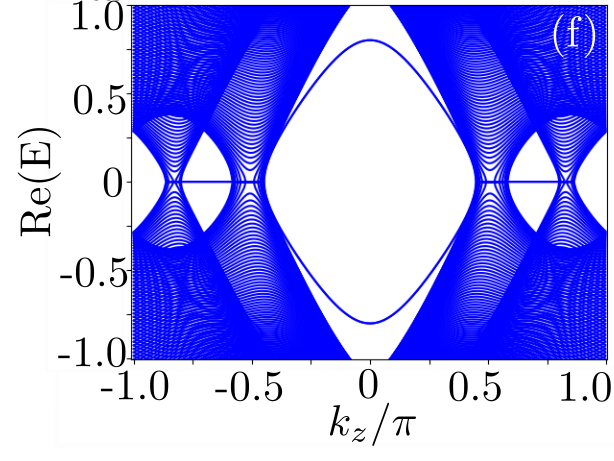} \\ 
    \includegraphics[width=0.28\linewidth]{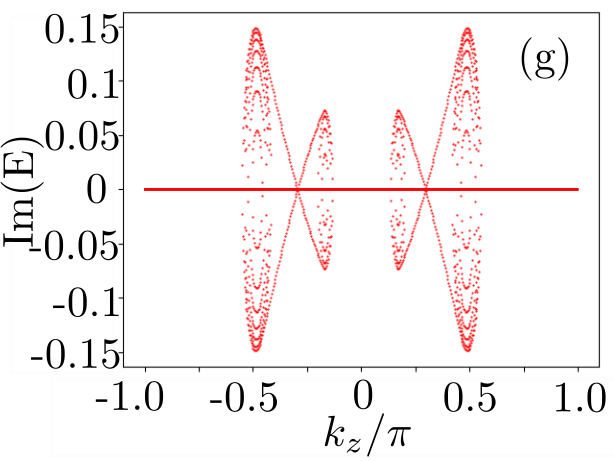} & \hspace*{-.1cm}\includegraphics[width=0.28\linewidth]{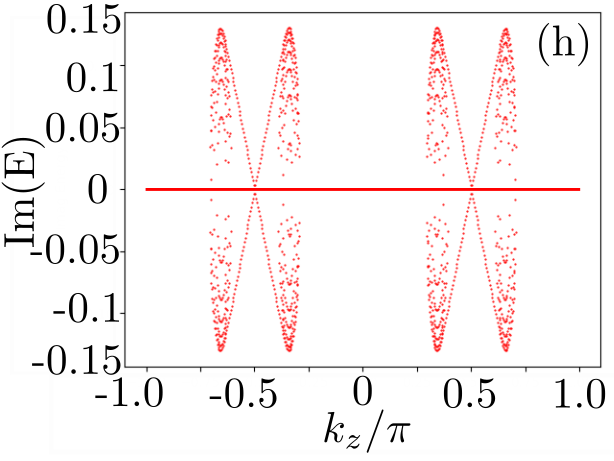} &
    \hspace*{-.1cm} \includegraphics[width=0.28\linewidth]{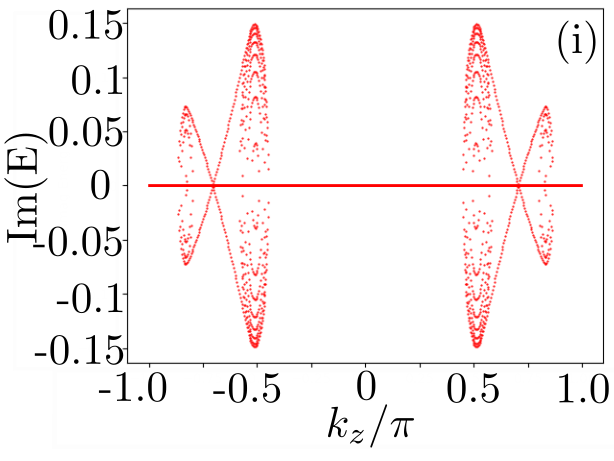} \\ 
    \includegraphics[width=0.28\linewidth]{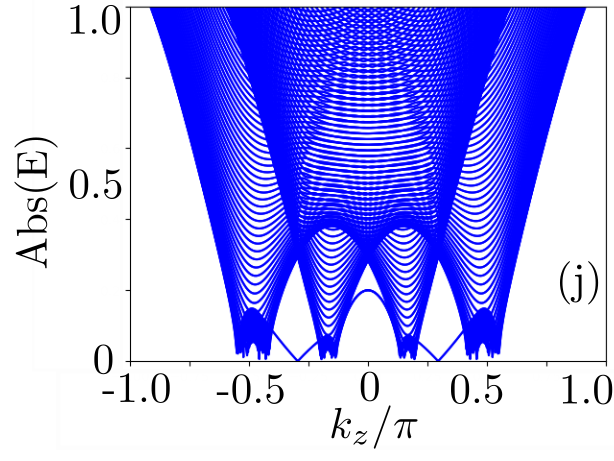}&
    \hspace*{-.1cm}\includegraphics[width=0.28\linewidth]{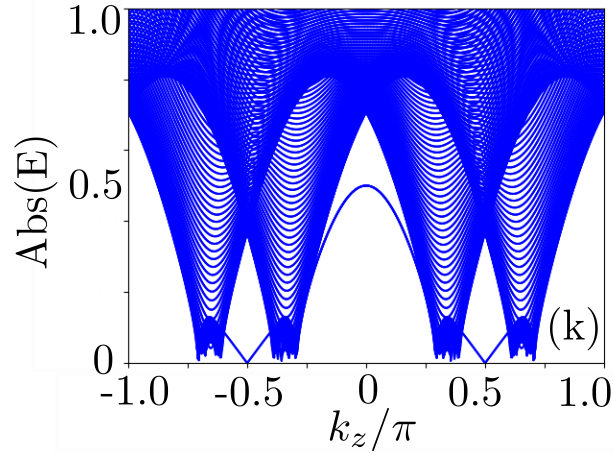} & 
    \hspace*{-.1cm}\includegraphics[width=0.28\linewidth]{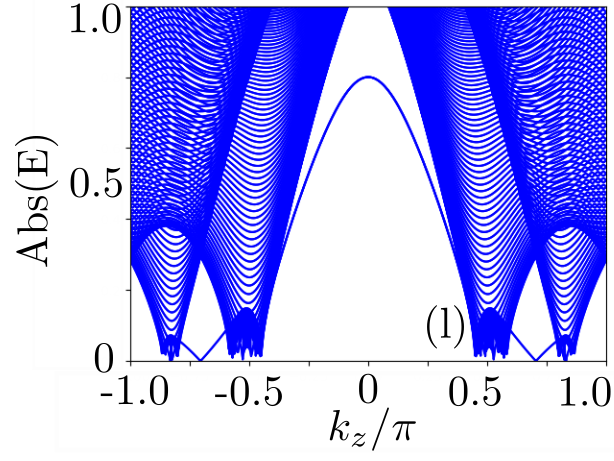} \\
          \includegraphics[width=0.28\linewidth]{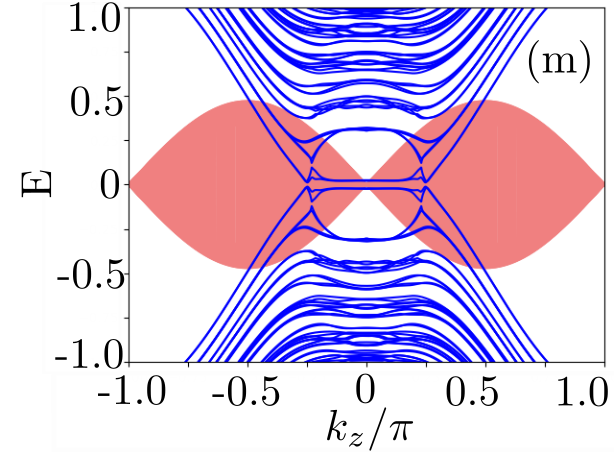}&
          \hspace*{-.1cm}\includegraphics[width=0.28\linewidth]{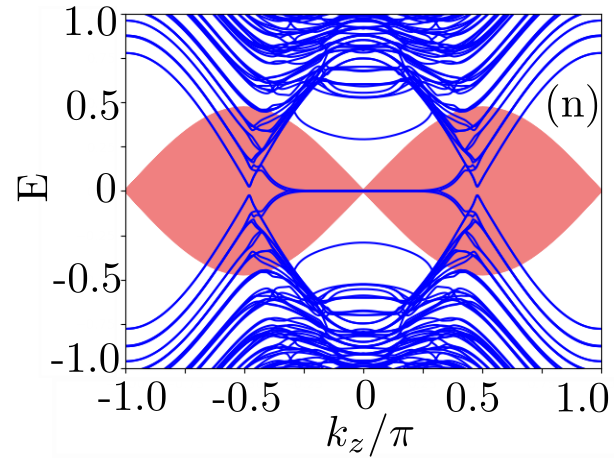}&
          \hspace*{-.1cm}\includegraphics[width=0.28\linewidth]{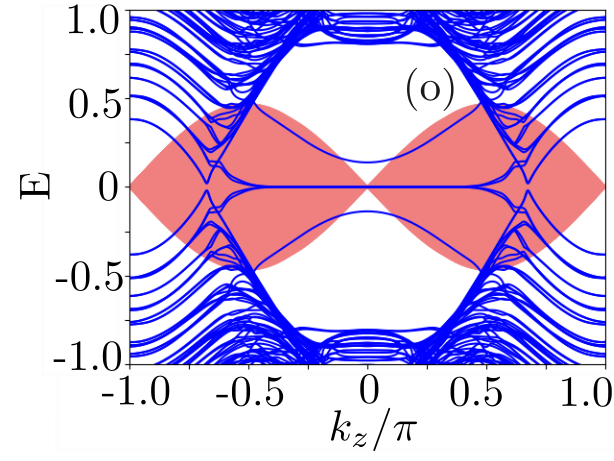} \\ \includegraphics[width=0.28\linewidth]{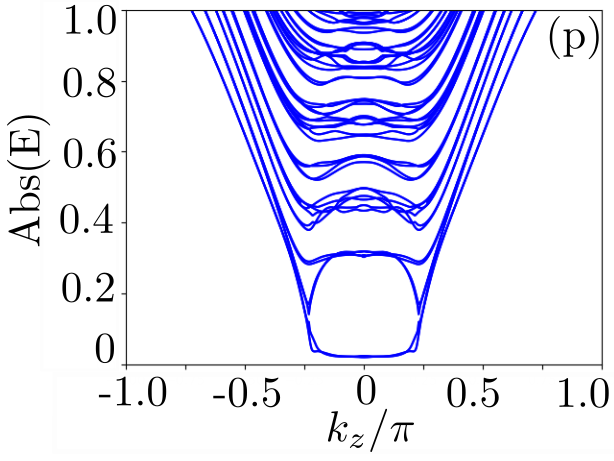} &  \hspace*{-.1cm}\includegraphics[width=0.28\linewidth]{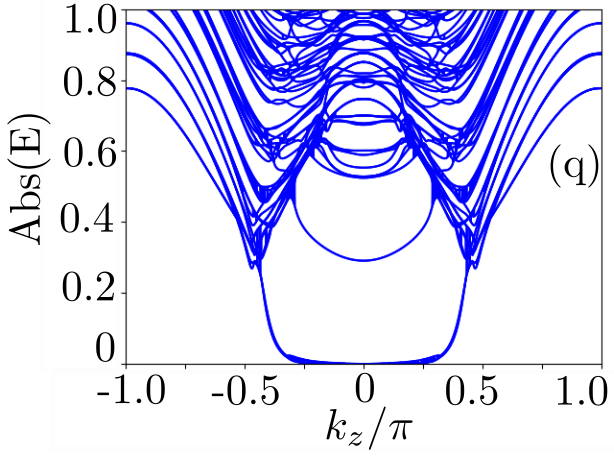}&
          \hspace*{-.5cm}\includegraphics[width=0.28\linewidth]{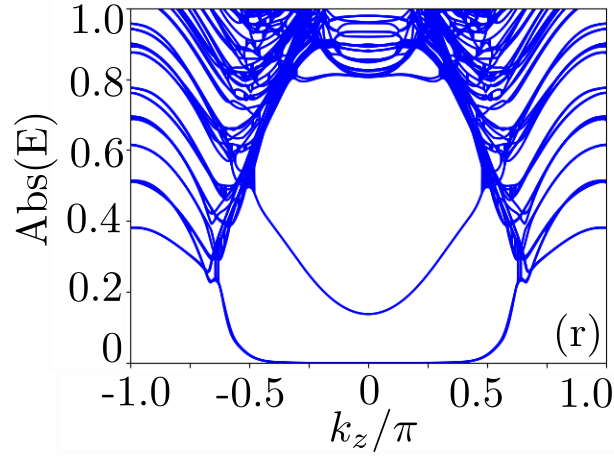}
\end{tabular}
   
    \caption{(a)-(c) Bulk spectrum for the time-reversal symmetric model written in Eq. (\ref{HTR}), for $\gamma=(-1.3,-1.0,-0.7)$. (d)-(f) Corresponding real eigenspectrum of $x$ surface ($k_y=0$) along $k_z$, showing surface FAs. (g)-(i) Imaginary eigenspectrum along $k_z$ (j)-(l) Same $x$ surface but with absolute energy depicting DPs with modulating dispersion with change in $\gamma$. (m)-(o) Hinge spectrum with two-axis OBC for inversion symmetric model. (p)-(r) Corresponding hinge absolute energy spectrum. Other parameters are the same as used in Fig. \ref{invsym}.}
    \label{trsym}
\end{figure}

Unlike the  inversion symmetric model written in Eq. (\ref{IH}), here we introduce momentum dependence to the Weyl perturbation along with NH perturbation. Due to this change, the model breaks inversion symmetry ($\mathcal{P}$) but in turn, preserves time-reversal symmetry ($\mathcal{T}$) with complex conjugate ($\mathcal{K}$) being the time reversal operator.
\begin{equation}
    H^{\mathcal{T}}_{\text{NHW}}(k)=H'_0(k)+i\, \delta\sigma_0\kappa_1sin(k_z),
    \label{HTR}
\end{equation}
where $H'_0(k)$ is the Hermitian part with momentum dependant Weyl perturbation,
\begin{align}
    &H'_0(k)=  \sum_{j=1}^4 h_j \Gamma_j +  m \sigma_0\kappa_2 sin(kz).
\end{align}
The band diagrams of the TR symmetric model corresponding to the bulk, one-axis OBC surface, and two-axis OBC hinges are plotted in Fig. \ref{trsym}. In contrast to the inversion symmetric model, this model hosts eight EPs in the bulk for all three values of $\gamma$. We notice that out of eight EPs, four are of the normal order in nature as they are connected by the surface states. At $\gamma=-1.3$, hinge states do not appear, and thus, higher-order EPs are not present. As we vary $\gamma$, two EPs closest to $k_z=0$ transform into higher-order and are connected by four-fold degenerate hinge states, along with the remaining four first-order EPs.  \\
Figs. \ref{invsym} and \ref{trsym} highlight the fact that there is no significant deviation  in the outcome from the model discussed in the main text. This implies that these symmetries ($\mathcal{I}$ and $\mathcal{T}$) turn out to be futile for the model in the main text. Another interesting point is, the momenta values of DP and the form of low energy dispersion around them do not change when we are at Dirac phase $(\gamma=-1)$ or in the Luttinger phase $(\gamma=-0.5)$ even after invoking the above symmetries. We, therefore, conclude that the DPs on the surface do not require these symmetries to stabilize them.

\end{document}